\documentclass[review]{elsarticle}
\usepackage{lineno}
\usepackage{xcolor}
\usepackage[hidelinks]{hyperref}
\usepackage{amsmath}
\usepackage{amsthm}
\usepackage{amsfonts}
\usepackage{amssymb}
\usepackage{graphicx}
\usepackage{soul,color}
\journal{Journal of \LaTeX\ Templates}
\usepackage{geometry}

\newtheorem{thm}{Theorem}

\theoremstyle{definition}

\newproof{pf}{Proof}
\newproof{pot}{Proof of Theorem \ref{thm2}}

\newtheorem{lem}{Lemma}

\newtheorem{prop}{Proposition}

\theoremstyle{definition}
\newtheorem{defn}{Definition}

\newtheorem{rem}{Remark}

\modulolinenumbers[5]

\begin{document}
	
	\begin{frontmatter}
		
		%\title{Positive T--S Fuzzy Controller Design for Cancer Tumor-Immune System under Immunotherapy and Chemotherapy: A Dual Approach}
		%%%\title{A Dual Approach for Positive T--S Fuzzy Controller Design and Its Application to Cancer Treatment under Immunotherapy and Chemotherapy}
		\title{A Dual Approach for Positive T--S Fuzzy Controller Design and Its Application to Cancer Treatment Under Immunotherapy and Chemotherapy}
		
		\author[add1]{Elham Ahmadi}
		\author[add1,add2]{Jafar Zarei\corref{mycorrespondingauthor}}
		\cortext[mycorrespondingauthor]{Corresponding author}
		\ead{zarei@sutech.ac.ir}
		\author[add2]{Roozbeh~Razavi-Far}
		\author[add2]{Mehrdad~Saif}
		%\author{Elham~Ahmadi, Jafar~Zarei, Roozbeh~Razavi-Far, Mehrdad~Saif}% <-this % stops a space
		%\ead{zarei@sutech.ac.ir}
		%
		%\author{\corref{mycorrespondingauthor}}
		%\ead{zarei@sutech.ac.ir}
		\address[add1]{Department of Electrical and Electronics Engineering, Shiraz University of Technology, Shiraz, Iran}
		\address[add2]{Department of Electrical and Computer Engineering, Windsor University, Windsor, ON, Canada}
		%\author[mysecondaryaddress]{Jafar Zarei\corref{mycorrespondingauthor}}
		\begin{abstract}
				This study proposes an effective positive control design strategy for cancer treatment by resorting to the combination of immunotherapy and chemotherapy. The treatment objective is to transfer the initial number of tumor cells and immune--competent cells from the malignant region into the region of benign growth where the immune system can inhibit tumor growth. In order to achieve this goal, a new modeling strategy is used that is based on Takagi--Sugen.
%				 \textcolor{red}{Based on Stepanova nonlinear model, a T--S fuzzy model, which allows designing effective controllers in a systematic way, is derived.}
				A Takagi-Sugeno fuzzy model is derived based on the Stepanova nonlinear model that enables a systematic design of the controller.
				Then, a positive Parallel Distributed Compensation controller is proposed based on a linear co-positive Lyapunov Function so that the tumor volume and administration of the chemotherapeutic and immunotherapeutic drugs is reduced, while the density of the immune-competent cells is reached to an acceptable level.
				Thanks to the proposed strategy, the entire control design is formulated as a Linear Programming problem, which can be solved very efficiently.
%			Using the proposed control strategy, the entire control design problem is recast \textcolor{red}{as Linear Programming (LP),} which can be solved very efficiently. 
			Finally, the simulation results show the effectiveness of the proposed control approach for the cancer treatment.
			%\textcolor{red}{Accordingly,} a nonlinear mathematical model which describes the dynamic relationship between tumor cells and the activity of the immune system throughout the development of cancer \textcolor{red}{is utilized.}
			%\textcolor{red}{It is worth mentioning that} this model is considered as a positive system, an approach that has not been studied in the literature. 
			%A situation that has not been considered by any of the existing design methods.
			%
			%The treatment objective is that moving an initial condition that lies in the malignant region into the region of benign growth of the tumor and therefore the cancer is controlled.
			%
			%%%The treatment objective is that an initial condition that is located in the malignant region is transferred to the region of benign growth, which in this region the immune system can inhibit tumor growth. 

			%To this end, first,
			%Firstly, a new Takagi--Sugeno (T--S) modeling approach is derived for this tumor-immune system interaction.
			% nonlinear mathematical model
			%Secondly,
			%To do this, a parallel distributed compensation (PDC) is proposed using linear co-positive Lyapunov function (LCPLF) and the obtained results are expressed in terms of linear programming (LP).
			%%%The proposed approach involves a new procedure for integrating the T--S fuzzy model, the linear co-positive Lyapunov function (LCPLF), and parallel distributed compensation (PDC).
			%
			%Finally,

		\end{abstract}
		
		
	\end{frontmatter}
	
	%\linenumbers
	
	\section{Introduction}
	
	%Cancer is a disease in which the cells lose their ability to divide and grow normally, and they can be proliferated uncontrolledly and can infect neighbor tissues.
	After cardiovascular disease, cancer is the most common cause of death in the world \cite{wang2009prognostic}.
	There are several therapies for the cancer treatment such as surgery, immunotherapy, radiotherapy, and chemotherapy.
	%It is worth mentioning that
	It is noticeable that even when the treatment is completed,  there is still a possibility of the disease.
	For this reason, many researchers have studied the cancer models, aiming to find a definitive cancer treatment \cite{chareyron2009model, teles2019cancer, cattani2008qualitative}.
	In the literature, chemotherapy is the most common way for the cancer treatment.
%	 one of the ways to treat cancer or it is temporary relief by using medicines or drugs.
	%
	%One of the ways to treat cancer or its temporary relief known as chemotherapy in the medical literature. Chemotherapy usually refers to the use of medicines or drugs to treat cancer.
	%
	%It’s often shortened to “chemo.” In this therapeutic approach, cells and in particular, microorganisms and cancer cells are destroyed by several chemical drugs that are used in a certain order or in certain combinations.
	In some cases, chemotherapy is the only treatment. However chemotherapy is mostly used along with other treatments, the main reasons for this are described as follows:
	i) cancer cells tend to grow fast, and chemotherapy drugs kill fast-growing cells. Since these drugs travel throughout the body, they can also affect normal and healthy cells that are fast-growing. Therefore, damage to healthy cells is the main side effect of the chemotherapy.
	ii) cancer cells are deforming in order to survive while under treatment, and for this reason, the chemotherapy treatment is often stopped.
	%%%After a while from the start of chemotherapy, cancer cells become deformed until do not destroy by medication and for this reason, chemotherapy is often stopped.
	In the last few decades, immunotherapy has become an important part of treating some types of cancers \cite{fister2005immunotherapy}.
	%%%This treatment uses the body's own immune system to help fight cancer and has fewer side effects compared to chemotherapy.
	This treatment uses the body's own immune system to cure cancer with  fewer side effects compared to the chemotherapy treatment.
	%It seems to work better when used with other types of treatment.
	%This can be done in a couple of ways: i. stimulating your own immune system to work harder or smarter to attack cancer cells and ii. giving you immune system components, such as man-made immune system proteins.
	%Immunotherapy works better for some types of cancer than for others. but for others it seems to work better when used with other types of treatment.
	As a result, the optimal way to blend manifold cancer treatments remains an open problem.
	
	Abundant efforts have been devoted to study the dynamics and interactions of the tumor and normal cells aiming to design appropriate control strategies, such that the tumor cells are eradicated as much as possible without harming the healthy cells. 
	%In 1998, a mathematical model has been presented for expressing the relationship between the tumor and immune system. This model specifies the dynamics between tumor cells, immune cells, and cytotoxic cells \cite{kirschner1998modeling}.
	Stepanova \cite{stepanova1979course} suggested a mathematical model for expressing the interactions between the tumor and immune system.  This model, in spite of its simplicity, and with a small number of parameters, plays an important role in resembling tumor-immune interactions.
	In \cite{d2012dynamics}, a metamodel for tumor-immune system interactions has been proposed, where the treatment has been formulated as an optimal control problem by cytotoxic agents and immunostimulations.
	In \cite{rocha2018multiobjective}, a multi-objective optimal control of a tumor growth model with the immune response and drug therapies has been presented, so that the average number of tumor cells and immuno and chemotherapeutic drugs administration are minimized at the same time.
	%%%A comparison between the dynamics of three tumor growth models that consist of an immune system and a drug administration therapy by applying optimal control has been investigated in \cite{martins2016comparing}.
	In \cite{martins2016comparing}, a comparison has been made by means of an optimal strategy to control the dynamics of three tumor growth models that consist of an immune system and drug administration therapy.
	
	Up to now, most of the control strategies are based on minimizing the drug dosage in an open-loop mode using an optimal control approach.
	%Some of these control structures are in an open-loop mode that increases the sensitivity of the result to uncertainties and errors.
	Model predictive control (MPC) has been applied as an effective strategy for the cancer chemotherapy \cite{chen2012optimal}. Blended immunotherapy and chemotherapy for tumor treatment using MPC has been proposed in \cite{chareyron2009mixed}.
	The Kirschner model has been proposed in \cite{kirschner1998modeling}, and adopted to design an adaptive fuzzy back-stepping control strategy for the tumor-immune system in \cite{nasiri2018adaptive}. The purpose of this work is to develop an effective treatment plan for the drug dosage to reduce the volume of tumor cells. However, this case only covers cancer immunotherapy.
	
	In the past two decades, significant advances have been made for the control of nonlinear systems. Takagi--Sugeno (T--S) fuzzy models have widely used  in the control community to design nonlinear control strategies \cite{takagi1985fuzzy, tanaka2004fuzzy, benzaouia2016advanced}. This model consists of a convex combination of several linear subsystems and nonlinear membership functions in the form of IF-THEN rules, which are considered as a universal approximator for nonlinear dynamics.
	
	%Since any model of cancer involve quantities that have intrinsically
	%constant sign, i.e., they are always positive.
	This work considers the reformed Stepanova model for the cancer-immune system interactions.
	%under mixed immunotherapy and chemotherapy
	Although this model is a simple and basic model, it includes many important medical features.
	%But here, the Gompertz model is considered for cancer growth instead of the exponential model utilized in the basic formulation.
	%It is worth noting that 
	The state variables in this model involve quantities that have  non-negative signs.  This model belongs to a significant class of systems known as ``Positive Systems". In such systems, the state variables that correspond to any non-negative initial states always have to be non-negative \cite{farina2011positive, kaczorek2012positive}.
	In addition, the positive systems states are defined inside a cone situated in the positive orthant instead of the whole space.
	Such systems have several applications in controlling chemical processes, biology, medicine, and economics systems \cite{cacace2017positive}.
	%\textcolor{pink}{The stability of positive systems are evaluated by two different approaches.
	%The first approach utilizes a quadratic Lyapunov function (QLF) with the main difference that the Lyapunov function imposed to be diagonal and the obtained results are expressed in the linear matrix inequalities (LMIs) framework.
	%%due to the positiveness characteristic of the underlying system.
	%%The results of this approach can be written in the linear matrix inequalities (LMIs) framework.
	%The second approach emphasizes the states of positive systems that are nonnegative and uses a linear co-positive Lyapunov function (LCPLF). The results of this approach are based on linear programming (LP) formulation, which is simpler than LMI and results in less computational complexity.}
	
		This paper is a mixture of immunotherapy and chemotherapy for the cancer treatment.
		The purpose of this research is to design a positive T--S fuzzy controller for an appropriate and efficient treatment, in which the combination of tumor volume as well as the potential side effects are reduced, and, consequently, the density of immune--competent cells is reached to an admissible level.
		Simulation results show the effectiveness of the proposed method, which can help to develop mixed therapies.
		Compared to the state-of-the-art approaches, it is observed that during this treatment the maximum number of cancer cells and the side effects are significantly reduced.
	
	%In this paper, the mixed immunotherapy, and chemotherapy are used for cancer treatment.
	%%%\textcolor{blue}{According to the fact that the fuzzy model exactly represents the original nonlinear systems}
	The main contributions of this paper can be summarized as follows: a new fuzzy modeling framework, which exactly represents the nonlinear system, is proposed, which reduces the number of fuzzy rules that inherently affect the conservatism.
	Then, a parallel distributed compensation (PDC) is derived as a naive and reliable method to handle nonlinear control systems. 
	%Moreover, this controller does not have the problem of the lack of continuity that exists in the piecewise state-feedback controllers.
	%----->In this technique, the concept of dual system is utilized.
	The concept of a dual system is also employed.
	%using linear co-positive Lyapunov function (LCPLF).
	The previous methods to stabilize the positive systems use a quadratic Lyapunov function that is imposed to be diagonal, and, consequently, increases conservatism. Here, a linear co-positive Lyapunov function (LCPLF) is proposed. The obtained results are expressed in terms of linear programming (LP) that leads to  more convenient conditions for computation and analysis compared to the existing ones.
%	 The results of this approach lead to more convenient conditions for computation and analysis compared to the existing ones.
	However, the fuzzy model cannot be easily extracted from the reformed Stepanova model due to the existence of a constant term in the nonlinear model. 
	%Regarding the existence of a constant term in the reformed Stepanova model, the fuzzy model cannot be easily extracted from this nonlinear model.
	To cope with this problem, this approach considers the control input has two parts;
	%\textcolor{green}{OR, In order to extract a fuzzy model directly from the nonlinear equations and also reduce the conservatism due to the reduction of fuzzy rules, the control input is suggested so that}
	one part of this control agent eliminates some nonlinear terms, and the other part is responsible for the stability of the remaining nonlinear system.
	%In this approach, the control input has two parts;
	%
	%
	%The purpose of this paper is to illustrate that the positive T--S fuzzy controller design is an appropriate and efficient treatment method for cancer therapy under mixed immunotherapy and chemotherapy of tumor.
	%This approach is based on the Lyapunov based method, aiming to reduce tumor volume and also the potential side effects of treatment as well as the density of immune competent cells reaches an acceptable level.

		The rest of the paper is organized as follows: 
		The Stepanova model of the tumor-immune system is described in Section 2. 
		%Section 3 presents, very briefly, the theories of T--S fuzzy systems and positive systems.
		The proposed fuzzy modeling framework and the proposed control method for the positive T--S system are presented in Section 3.
		%%itive system. Section 4 presents 
		%%% ----->and provides asymptotic stability and stabilization conditions for the T--S positive system under LP framework.
		The simulation results are presented in Section 4. Section 5 concludes the paper.
		%%%------> \textcolor{green}{Moreover, the theories of T--S fuzzy systems and positive systems are briefly discussed in Appendix section.}
	
	\textbf{Notations:} $\mathbb{R}^n $ represents a set of \textit{n}-column real vectors, and $ \mathbb{R}^{n\times m} $ is a set of $n\times m$ real matrix. The notation $ A\geq\geq B $ (respectively, $A\gg B$) means that the matrix $A-B$ is element-wise nonnegative (respectively, $A-B$ is element-wise positive). $\textbf{1}=[1,1,...,1]^T$ denotes a column vector, in which all entries are equal to 1. The superscript `$T$' stands for the transpose operator; $I$ and $0$ represent the identity and zero matrices, respectively; diag$(A_1,...,A_n)$ represents a block diagonal matrix, in which $A_1,...,A_n$ are square matrices along the main diagonal. 
	
		\section{Stepanova Model}
		%\section{Preliminary results}
		In this section, the Stepanova mathematical model that expresses the interactions between the immune and tumor system is introduced \cite{stepanova1979course}. 
		%Next, a brief review of T--S fuzzy systems and positive systems is provided.
		%Moreover, several observations and results associated with the positive T--S fuzzy systems are introduced that will be utilized throughout this paper.
		%
		%\subsection{Stepanova model}
		%In this section,
		Here the Stepanova model is used to describe the tumor growth with an immune response, immunotherapy, and chemotherapy.
		To clarify this subject, one can consider the reformed form of the Stepanova model as follows \cite{d2012dynamics}:
		\begin{align}
		\dot{x}_1(t)&=\mu_cx_1(t) F(x_1) -\gamma x_1(t)x_2(t)-k_{x_1}x_1(t)u_1(t) \label{eq1} \\
		\dot{x}_2(t)&=\mu_I(x_1(t)-\beta x_1^2(t))x_2(t)-\delta x_2(t) + \alpha +k_{x_2}x_2(t)u_2(t) \label{eq2}
		\end{align}
		where $x_1(t) $ and $x_2(t) $ indicate the number of tumor cells and effector cells of the immune system, respectively. The control agent $u_1(t) $ denotes the blood profiles of a cytotoxic agent and $u_2(t) $ indicates the immune system booster drug for the immune cells in the cancer treatment. $k_{x_1}$ and $k_{x_2}$ stand for the maximum dose rates of the therapeutic agents.
	
	    Equation \eqref{eq1} models tumor growth, where $\mu_c$ denotes the tumor growth coefficient and the coefficient $\gamma$ shows the rate at which
		cancer cells are eliminated. The term $\gamma x_1x_2$ models the useful effect of the immune response on the cancer volume. $F$ is a function of the growth of the tumor cells and
		there are various selections that can be considered for that. However, here the Gompertz growth function \cite{norton1988gompertzian} in the form of $F(x_1) = - ln(\frac{x_1(t)}{x_\infty})$ is considered instead of other growth functions such as exponential model or logistic or generalized logistic growth. In this function, $x_\infty$ is the fixed finite carrying capacity of the tumor.
	%------->Lastly, $k_{x_1}$ is chemotherapeutic killing parameter.
	
	  Equation \eqref{eq2} models the main specifications of the immune system’s response to the tumor cells. The effect of the tumor growth on the activity of immune cells is shown by the first term. 
		The coefficients $\mu_I$, $\beta$ and $\delta$ are the tumor stimulated proliferation rate, the inverse threshold for tumor suppression and the rate of natural death of immune cells, respectively. $\alpha$ represents the rate of influx of the effector cells from the primary organs.
	%----------->Lastly, $k_{x_2}$ is chemotherapeutic killing parameter.
	
	%$\alpha$ models the constant rate of generation of T-cells as the most important factors to fight tumor growth.
	%
	%
	%The description of parameters in this model and their values are presented in Table \ref{tab_1}.
	%
	
	\textbf{Equilibrium point:}
	The determination of the equilibrium points and analyzing systems behavior around these points play a significant role in dynamical systems.
	Stopping the growth and moving process in these points are some of the most important reasons behind this subject.
	Although this is not suitable for some physical processes such as motion, it can be an ideal solution  for some medical applications, such as tumor growth \cite{merola2008insight}.
	%In studying a dynamical system, it is important to determine the equilibrium point of the system and analyze its behavior around the equilibrium point. 
	%One of the most important reasons is to stop the process of growth and movement in reaching these points, perhaps this is not a pleasure to examine some physical processes, such as motion, but it can be ideal for some medical issues, such as tumor growth.
	
	When the patient's body system reaches a point, in which the number of tumor cells is zero or very low, the treatment is completed. The unforced system without medication has two locally asymptotically stable equilibriums, which can be obtained by solving the equations $\dot{x}_1(t)=0$ and $\dot{x}_2(t)=0$, one microscopic point at (72.961, 1.32), which is the region of attraction corresponding to the benign conditions (where the number of tumor cells is low) and the other is macroscopic point at (737.278, 0.032), which is the region of attraction related to the malignant situations, where the number of tumor cells is high. These regions of attraction are separated by a stable manifold of an intermediate saddle point at (356.2, 0.439).
	
	It is obvious that for any model of cancer, the states $x_1(t)$ and $x_2(t)$ should remain positive for the positive initial conditions $ {x_1}(0) $ and $ {x_{2}(0)}$, and arbitrary acceptable controls $u_1(t)$ and $u_2(t)$. Therefore, the nonlinear model mentioned for cancer, based on the interaction between the tumor cells and the immune cells, can be considered as a positive system. Therefore, a set $\kappa$ that can be defined as follow is a  positive invariant set for the above model, and, then, no positive constraints imposed to variables:
	\begin{equation*}
	\kappa = \left\lbrace (x_1,x_2): x_1>0, x_2>0 \right\rbrace.
	\end{equation*}

		\section{Positive T-S Fuzzy Modeling and Controller design }
		In this section, first, a new fuzzy modeling framework for the nonlinear model of the cancer treatment is proposed so that the number of fuzzy rules is reduced. 
		Next, the stability and new LP conditions for the controller design of positive T--S systems are proposed.

		\subsection{Fuzzy Modeling}
		%As observed, before the system equations are nonlinear.
		To obtain a T--S fuzzy representation of Eqs. \eqref{eq1} and \eqref{eq2}, the nonlinear terms of the model are considered as premise variables. Denote $\vartheta_1=-\mu_c ln(\dfrac{x_1}{x_\infty}) $, $ \vartheta_2=-\gamma x_1 $, $\vartheta_3=\mu _I (x_1 - \beta x_1^2)-\delta $, $ \vartheta_4= - k_{x_1} x_1 $ and, $ \vartheta_5= k_{x_2} x_2 $. Therefore, the corresponding model is given by:
	\begin{align}
\left\lbrace\begin{array}{l}
\dot{x}_1 = \vartheta_1 x_1 - \vartheta_2 x_2 - \vartheta_4 u_1\\ 
\dot{x}_2 = \vartheta_3 x_2 + \alpha + \vartheta_5 u_2.
\end{array}  \right. 
	\label{a}
	\end{align} 
	Because of the existence of a constant term $\alpha$ in the second equation, the fuzzy model cannot be easily extracted from the system \eqref{a}.
	To overcome this problem, an approach is suggested, which is based on the change of variable.
	For this purpose, the control input consists of two parts:
	i) one part is used to eliminate some nonlinear terms of equations;
	ii) another part is considered to stabilize the reminding nonlinear dynamical model.
	%Then, the direct Lyapunov method is used for stability analysis and stabilizing controller design.
	%i. Part of the control agents is to eliminate some nonlinear terms of equations.\\
	%ii. Part of it is also responsible for the stability of the remaining nonlinear system.
	This approach has two main advantages; First, a T--S fuzzy model can be extracted directly from nonlinear equations of the system, and secondly, the number of fuzzy rules is reduced that leads to less conservatism.
	Accordingly, the new control input vector is defined as follows:
	\begin{equation}
	u_2^*=\alpha + k_{x_2} x_2 u_2 \ \rightarrow \ u_2 = \dfrac{u_2^* - \alpha}{k_{x_2} x_2}
	\label{b}
	\end{equation}
	Therefore, the equations \eqref{eq1} and \eqref{eq2} can be rewritten as follows:
	\begin{equation}
	\begin{cases}
	\dot{x}_1= - \mu_c x_1 ln(\dfrac{x_1}{x_\infty}) - \gamma x_1 x_2 - k_{x_1} x_1 u_1,\\
	\dot{x}_2= \mu_I (x_1 - \beta x_1^2)x_2 - \delta x_2 + u_2^*.
	\label{c}
	\end{cases}
	\end{equation} 
	By defining the nonlinear terms as $\bar{\vartheta_1}=-\mu_c ln(\dfrac{x}{x_\infty})- \gamma y $, $\bar{\vartheta_2}=-\mu_I(x-\beta x^2)- \delta $ and,  $\bar{\vartheta_3}=-k_x x $, then the general form of the Eq.\eqref{c} is as follows:
	\begin{equation}
	\begin{cases}
	\dot{X}(t)=A(\bar{\vartheta}_1,\bar{\vartheta}_2) X(t) + B(\bar{\vartheta}_3)U(t),\\
	\ z(t)=CX(t),
	\label{d}
	\end{cases}
	\end{equation}
	where $z(t)$ is the output vector and we have,\\
	\begin{align*}
	X=\begin{bmatrix}
	x_1 \\ 
	x_2
	\end{bmatrix}, \ \
	A(\bar{\vartheta}_1,\bar{\vartheta}_2)=\begin{bmatrix}
	\bar{\vartheta}_1 & 0 \\ 
	0 & \bar{\vartheta}_2
	\end{bmatrix}, \ \
	B(\bar{\vartheta}_3)=\begin{bmatrix}
	\bar{\vartheta}_3 & 0 \\ 
	0 & 1
	\end{bmatrix},\ \
	U=\begin{bmatrix}
	u_1 \\ 
	u_2^*
	\end{bmatrix},\ \
	C=\it I_2.  
	\end{align*}
	%In this cases, the tracking problem is considered, accordingly, the following integrator is used:
	In the set point tracking problem, the following integral control structure is used:
	\begin{equation*}
	e_I= \int (z_r-z)dt
	\end{equation*}
	where $e=z_r-z$ is the tracking error. The goal is that the output $z$ tracks a given reference $z_r$. Finally, the augmented system can be defined by the following system:
	\begin{equation}
	\begin{cases}
	\dot{\bar{X}}(t)=\bar{A}(\bar{\vartheta}_1,\bar{\vartheta}_2)\bar{X}+\bar{B}(\bar{\vartheta}_3)U + \bar{D} Z_r,\\
	\ z(t) = \bar{C} \bar{X}.
	\end{cases}
	\end{equation}
	where
	\begin{align*}
	\bar{X}=\begin{bmatrix}
	X \\ 
	e_I
	\end{bmatrix}&,\ \
	\bar{A}(\bar{\vartheta}_1,\bar{\vartheta}_2)=\begin{bmatrix}
	A(\bar{\vartheta}_1,\bar{\vartheta}_2) & 0 \\ 
	-C & 0
	\end{bmatrix}, \ \
	\bar{B}(\bar{\vartheta}_3)=\begin{bmatrix}
	B(\bar{\vartheta}_3) \\ 
	0
	\end{bmatrix},\\
	&\bar{C}=\begin{bmatrix}
	C & 0
	\end{bmatrix},\ \
	\bar{D}=\begin{bmatrix}
	0 & 0 \\ 
	0 & I
	\end{bmatrix},\ \
	Z_r=\begin{bmatrix}
	0 \\ 
	z_r
	\end{bmatrix} .    
	\end{align*}

	By computing the maximum and minimum values of premise variables under $x_1 \in [0,1000] $ and $ x_2 \in[0,5] $, $\bar{\vartheta_1}$, $\bar{\vartheta_2}$ and, $\bar{\vartheta_3}$ can be described as:
	%%%The maximum and minimum values of $\bar{\vartheta_1}$, $\bar{\vartheta_2}$ and, $\bar{\vartheta_3}$ under $x_1 \in [0,1000] $ and $ x_2 \in[0,5] $ are computed as follows:
	%%%\begin{align*}
	%%%&\text{max}(\bar{\vartheta_1})=5.0178, \qquad \ \  \text{min} (\bar{\vartheta_1})=-5.1391, \\
	%%%&\text{max} (\bar{\vartheta_2})=-0.3740, \qquad \text{min} (\bar{\vartheta_2})=-8.3121,\\
	%%%&\text{max} (\bar{\vartheta_3})=-0.1000, \qquad \text{min} (\bar{\vartheta_3})=-1000,
	%%%\end{align*}
	%
	%%%Therefore, $\bar{\vartheta_1}$, $\bar{\vartheta_2}$ and, $\bar{\vartheta_3}$ can be described as:
	\begin{align*}
	\bar{\vartheta}_1 &= \text{max}(\bar{\vartheta_1}) \ . \ M_1(\bar{\vartheta}_1)+ \text{min} (\bar{\vartheta_1}) \ .  \ M_2(\bar{\vartheta}_1),\\
	\bar{\vartheta}_2 &= \text{max} (\bar{\vartheta_2}) \ . \ N_1(\bar{\vartheta}_2)+ \text{min} (\bar{\vartheta_2}) \ .  \ N_2(\bar{\vartheta}_2),\\
	\bar{\vartheta}_3 &= \text{max} (\bar{\vartheta_3}) \ . \ S_1(\bar{\vartheta}_3)+ \text{min} (\bar{\vartheta_3}) \ .  \ S_2(\bar{\vartheta}_3),
	\end{align*}
	where
	% $M_1(\bar{\vartheta_1})=\dfrac{\bar{\vartheta_1}-\text{min} (\bar{\vartheta_1})}{\text{max}(\bar{\vartheta_1})-\text{min} (\bar{\vartheta_1})}$, $M_2(\bar{\vartheta_1})=1-M_1(\bar{\vartheta_1})$, $N_1(\bar{\vartheta_2})=\dfrac{\bar{\vartheta_2}-\text{min} (\bar{\vartheta_2})}{\text{max} (\bar{\vartheta_2})-\text{min} (\bar{\vartheta_2})}$, $N_2(\bar{\vartheta_2})=1-N_1(\bar{\vartheta_2})$,
	%$S_1(\bar{\vartheta_3})=\dfrac{\bar{\vartheta_3}-\text{min} (\bar{\vartheta_3})}{\text{max} (\bar{\vartheta_3})-\text{min} (\bar{\vartheta_3})}$ and, $S_2(\bar{\vartheta_3})=1-S_1(\bar{\vartheta_3})$.
	\begin{align*}
	& M_1(\bar{\vartheta_1})=\dfrac{\bar{\vartheta_1}-\text{min} (\bar{\vartheta_1})}{\text{max}(\bar{\vartheta_1})-\text{min} (\bar{\vartheta_1})},
	M_2(\bar{\vartheta_1})=1-M_1(\bar{\vartheta_1})\\
	& N_1(\bar{\vartheta_2})=\dfrac{\bar{\vartheta_2}-\text{min} (\bar{\vartheta_2})}{\text{max} (\bar{\vartheta_2})-\text{min} (\bar{\vartheta_2})},
	N_2(\bar{\vartheta_2})=1-N_1(\bar{\vartheta_2}),\\
	& S_1(\bar{\vartheta_3})=\dfrac{\bar{\vartheta_3}-\text{min} (\bar{\vartheta_3})}{\text{max} (\bar{\vartheta_3})-\text{min} (\bar{\vartheta_3})},
	S_2(\bar{\vartheta_3})=1-S_1(\bar{\vartheta_3}).
	\end{align*}
	Accordingly, the membership functions are calculated as follows:
	\begin{align*}
	h_1 &= M_1 \times N_1 \times S_1, \quad h_2 = M_1 \times N_1 \times S_2,\\
	%h_2 &= M_1 \times N_1 \times S_2,\\
	h_3 &= M_1 \times N_2 \times S_1, \quad h_4 = M_1 \times N_2 \times S_2,\\
	%h_4 &= M_1 \times N_2 \times S_2,\\
	h_5 &= M_2 \times N_1 \times S_1, \quad h_6 = M_2 \times N_1 \times S_2,\\
	%h_6 &= M_2 \times N_1 \times S_2,\\
	h_7 &= M_2 \times N_2 \times S_1, \quad h_8 = M_2 \times N_2 \times S_2.\\
	%h_8 &= M_2 \times N_2 \times S_2.
	\end{align*}
	
	It should be noted that the Euler technique is applied for discretization, however, any discretization method can also be used by a sampling time \textit{T}.
	Hence, the T--S fuzzy model related to the augmented discrete-time system is presented by 8 linear sub-models:
	%One can use any discretization method with a sampling time \textit{T}. However, to illustrate our method, only the Euler technique is used. Hence, the associated eight augmented discrete-time linear systems become:
	%
	\begin{equation}
	\begin{cases}
	&\bar{X}(k+1)=\bar{A}d_i \bar{X}(k) + \bar{B}d_i U(k) + \bar{D}d_i Z_r ,\\
	&z(k) = \bar{C} \bar{X}(k),
	\label{e}
	\end{cases}
	\end{equation}
	where $i=1,2,...,8$ and matrices $\bar{A}d_i$, $\bar{B}d_i$ and, $\bar{D}d_i$, respectively, characterize the discretized matrices related to $\bar{A}_i$, $\bar{B}_i$ and, $\bar{D}_i$.
	%
	%\begin{table}[h!]
	%\centering
	%\caption{Numerical values for the variables and parameters used in simulations \cite{d2012dynamics}.}
	%\label{tab_1}
	%\begin{tabular}{cccc}
	%\hline
	% Variable/Parameter & Description & Numerical value & Dimension\\
	% \hline
	% $x$ & Tumor volume & - & $10^6$ cells \\
	% $x_0$ & Initial value for $x$ & $600$ & $10^6$ cells\\
	% $y$ & Immune-competent cell density & - & Non-dimensional\\
	% $y_0$ & Initial value for $y$ &  0.1  &  Non-dimensional\\
	% $\alpha$ & Rate of influx & 0.1181 & 1/day\\
	% $\beta$ & inverse threshold  for tumor suppression & 0.00264 & Non-dimensional\\
	% $\gamma$ & Interaction rate & 1 & $10^7$ cells/day\\
	% $\delta$ & Death rate & 0.37451 & $1$/day\\
	% $\mu _C$ & Tumor growth parameter & 0.5599 &  $10^7$ cells/day\\
	% $\mu _I$ & Tumor stimulated proliferation rate & 0.00484 & $10^7$ cells/day\\
	% $x_\infty$ & Fixed carrying capacity & 780 & $10^6$ cells\\
	% $k$ & killing parameter & 1 &  $10^7$ cells/day\\
	% \hline
	%\end{tabular}
	%\end{table}
	%
	
	To validate the newly proposed fuzzy modeling framework, the nonlinear model is simultaneously simulated with the T--S fuzzy model of the cancer system.
	The parameters reported in the Table (\ref{tab_1}) are used for simulation.
	
\begin{figure}[h!]
		\begin{center}
			\includegraphics[scale = 0.5]{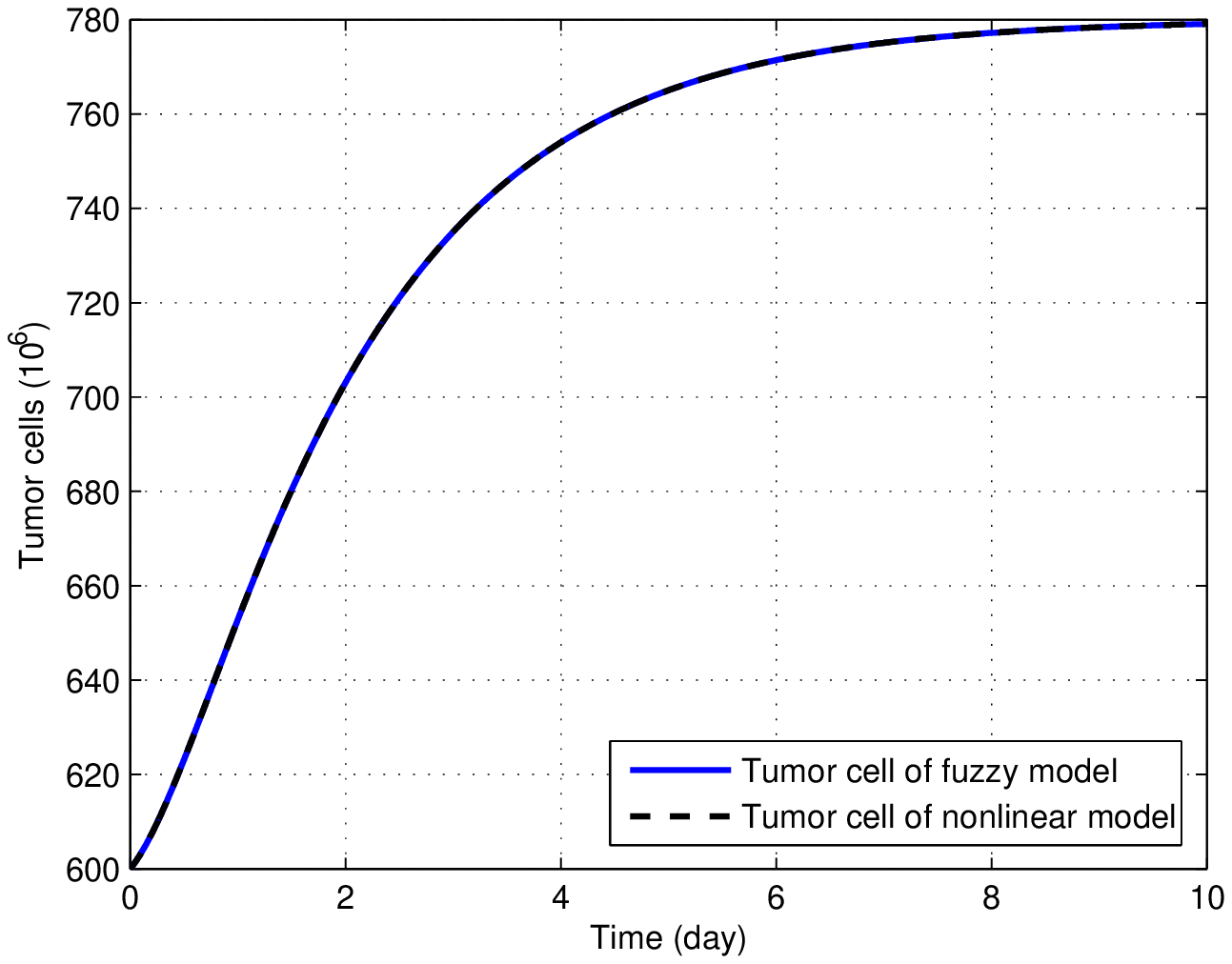}
			\includegraphics[scale = 0.5]{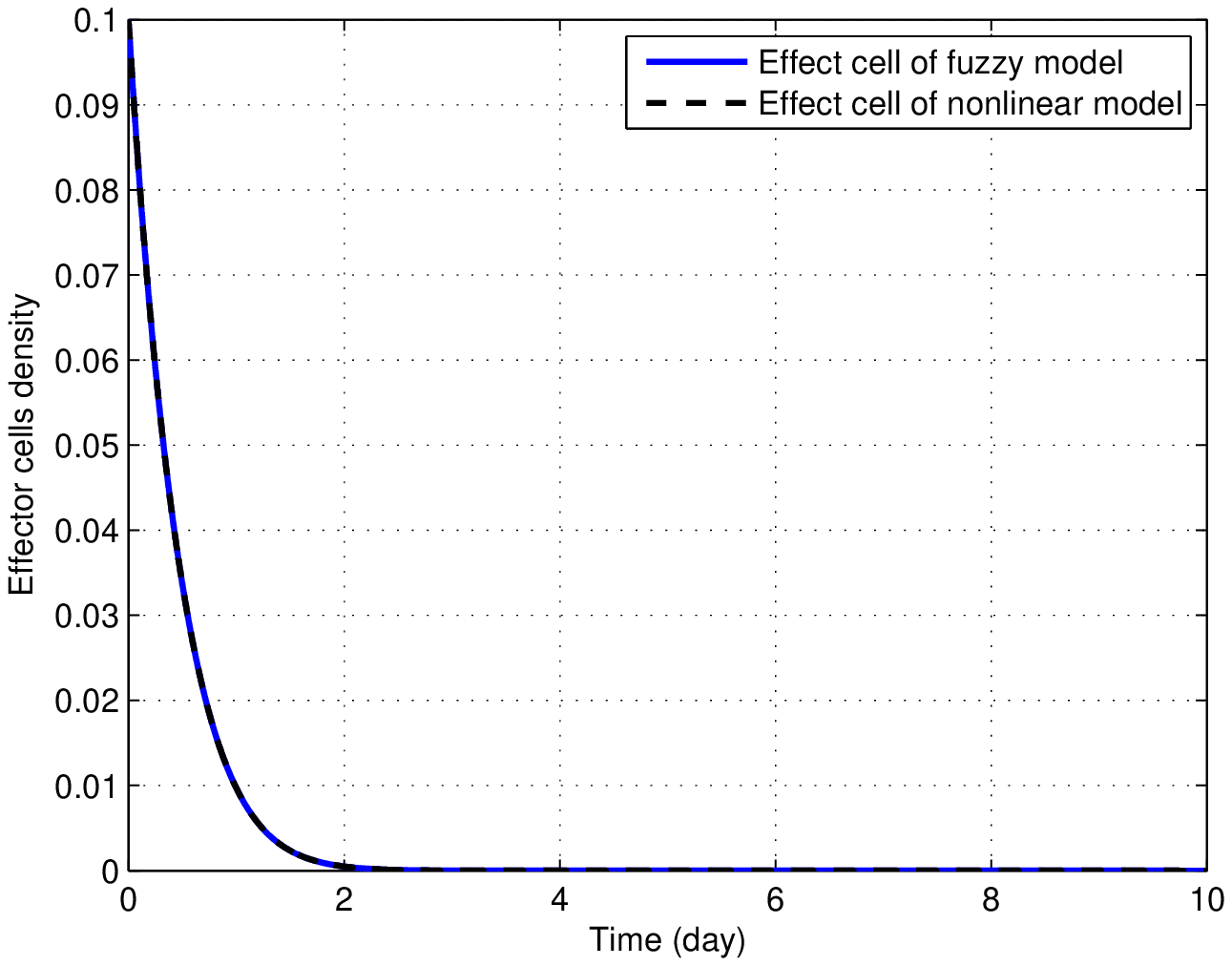}
			\caption{Dynamic profile of the tumor cells and the effector cells' density.}
			\label{figg1}
		\end{center}
	\end{figure}

	\begin{table}[h!]
		\centering
		\caption{ The values of parameters/variables \cite{d2012dynamics}.}
		\label{tab_1}
		\begin{tabular}{cccc}
			\hline
			Variable/Parameter & Descriptions & Numerical value & Dimension\\
			\hline
			% $x$ & Tumor volume & - & $10^6$ cells \\
			$x_1(0)$ & The initial value of $x_1$ & $600$ & $10^6$ cells\\
			% $y$ & Immune-competent cell density & - & Non-dimensional\\
			$x_2(0)$ & The initial value of $x_2$ &  0.1  &  Non-dimensional\\
			$\alpha$ & Rate of influx & 0.1181 & 1/day\\
			$\beta$ & Inverse threshold  for tumor suppression & 0.00264 & Non-dimensional\\
			$\gamma$ & Interaction rate & 1 & $10^7$ cells/day\\
			$\delta$ & Death rate & 0.37451 & $1$/day\\
			$\mu _C$ & Tumor growth parameter & 0.5599 &  $10^7$ cells/day\\
			$\mu _I$ & Tumor stimulated proliferation rate & 0.00484 & $10^7$ cells/day\\
			$x_\infty$ & Fixed carrying capacity & 780 & $10^6$ cells\\
			$k$ & chemotherapeutic killing parameter & 1 &  $10^7$ cells/day\\
			\hline
		\end{tabular}
	\end{table}
	%
	%The trajectories of the original nonlinear system and T--S model in the absence of control by taking initial condition as $(x_0,y_0)=(600,0.1)$ are shown in Fig. \ref{figg1}.
	The simulation results in the absence of control by taking an initial condition as $ X_0 = \begin{bmatrix}
	600 & 0.1
	\end{bmatrix} ^ T $, is shown in Fig. \ref{figg1}.
	%
	% $(x_0,y_0)=(600,0.1)$
	%
	The tumor volume is in multiples of $10^6$ cells, while the effector cells density is measured in a scale relative to 1.
	As can be seen, the behavior of both systems is quite similar and the T--S fuzzy system exactly models the nonlinear system.
	Fig. \ref{figg1} shows that simultaneously with the growth and proliferation of tumor cells, the density of the immune system cells decreases significantly.
	Therefore, these results substantiate the effectiveness of the synthesized fuzzy model that will be considered in the controller synthesis. 
	
	The goal is to design a controller that reduces the tumor volume and also holds the density of the effector cells at an acceptable level.
	%%%----> For this end, the controller must be able to move the initial condition that is located in the malignant region into the region of benign growth.
	Therefore, the controller is designed in a way that it guarantees the states $x_1$ and $x_2$ always remain positive.
	%%%---->, while the error \textit{e} is sign free.
	%Therefore, the positivity of the closed-loop system might be important to satisfy hard constraints on the sign of the states. 
	Then, the linear Lyapunov based method is used to design the controller that renders nonnegative states, while they do not always exist using the classical Lyapunov function even with linear matrix inequalities (LMIs).
	%Using the Lyapunov based method, we seek a controller that guarantees the existence of a control law that renders nonnegative states, while it does not always exist using the classical Lyapunov function even with LMIs.
	%This problem organize one of the main motivation of the present work which uses the mixed immunotherapy and chemotherapy of tumors for cancer treatment.

	%----->\section{Positive T--S fuzzy Control Strategies}
	
	%%%-------->This section includes two subsections. The first subsection represents
	%the stability condition for T--S positive systems without considering the control input. The next subsection proposes new LP conditions for the controller design so that the closed-loop system is asymptotically stable and positive.
	
		\subsection{Stability Analysis}
	The control objective of this subsection is to obtain the stability condition for the positive T--S fuzzy systems. The attained result is presented in the LP framework.
	Throughout this subsection it is assumed that the underlying system is unforced that is, $u(k)\equiv0$.
	\begin{rem} \label{last}
		A system whose matrices are transposed is called a dual system of another one. It is significant to note that asymptotic stability of the system \eqref{eq4} is equivalent to asymptotic stability of its dual \cite{ blanchini2015switched, benzaouia2018conditions}, i.e,
		\begin{equation}
		x(k+1) = \sum_{i=1}^{r} h_i(\vartheta(k)) A^T_i x(k).
		\label{eqqq5}
		\end{equation}
	\end{rem}
	
	\begin{prop}\label{prop1}
		If there exist a vector $\;p \geq \geq 0$ such that 
		\begin{equation}
		(A_i-I) p\ll0, \label{eq5b}
		\end{equation}
		holds $i=1,2,...,r$, then, the system \eqref{eq4} is asymptotically stable.
	\end{prop}
	%system \eqref{eq4} is asymptotically stable, if there exists a vector $\;p \geq \geq 0$ $(\; \bar{p} \geq \geq 0)$ satisfying one of the following two LP problems:
	%\begin{subequations}
	%\begin{align}
	%\text{LP}\ 1:& \; p^T(A_i-I)\ll 0, \label{eq5a}\\
	%\text{LP}\ 2:& \;\;(A_i-I)\bar{p}\ll0, \label{eq5b}
	%\end{align}
	%\end{subequations}
	%where $i=1,2,...,r.$}
	%\end{prop}
	
	\begin{pf}
		Given the positive system \eqref{eqqq5} means $x(k) \geq \geq 0$ for all $ k\in R^n_+$. Consider a linear Lyapunov function candidate as
		\begin{equation}
		V(x(k)) = x^T(k) p,\ \  p \geq \geq 0.
		\label{dual}
		\end{equation}
		Then the rate of increase of $V(x(k))$ is given by:
		%\begin{align} \label{eqq4}
		%\Delta V(x_k)&=p^Tx_{k+1}-p^Tx_k  \nonumber\\
		%&=\sum_{i=1}^{r}h_i(\vartheta(k))A_ix_k-p^Tx_k \nonumber\\
		%&=\sum_{i=1}^{r}h_i(\vartheta(k))p^T(A_i-I) x_k. 
		%\end{align}
		\begin{align}
		\Delta V(x(k))&=V(x(k+1))-V(x(k)),\nonumber\\
		&=x^T(k+1) p  - x^T(k) p,\nonumber\\
		&=\sum_{i=1}^{r}h_i(\vartheta(k))x^T(k)A_i p- x^T(k) p,\nonumber\\
		&=\sum_{i=1}^{r}h_i(\vartheta(k))\left(  x^T(k)(A_i-I) p \right) < 0,
		\label{eqqq6}
		\end{align}
		Since the states of positive systems are nonnegative and $0\leq h_i(\vartheta_k)\leq 1$ and $\sum_{i=1}^{r}h_i(\vartheta(k))=1$, the Lyapunov stability condition, $\Delta V(x(k))<0$ holds if and only if the inequality \eqref{eq5b} is satisfied.
		The whole proof is completed.  \qquad \qquad \qquad \qquad \qquad \qquad \qquad \qquad \qquad \qquad \qquad \qquad \qquad \qquad \qquad \ \ $\square $
		% 
		%%%%%Since the states of positive systems are nonnegative and $0\leq h_i(\vartheta_k)\leq 1$ and $\sum_{i=1}^{r}h_i(\vartheta(k))=1$
		%%%%%and from \eqref{eq5b}, $\Delta V(x(k))<0$ holds. The whole proof is completed.  \qquad \qquad \qquad \qquad \qquad \qquad \qquad  \ \ $\square $ 
		%Therefore system \eqref{eq4} is asymptotically stable.
		
		As a conclusion, the existence of LCPLF for the dual system \eqref{eqqq5} ensures the asymptotic stability of the original system \eqref{eq4}. 
		It should be noted that the existence of LCPLF for the positive system \eqref{eq4} neither guarantees the existence of such a Lyapunov function for the dual system \eqref{eqqq5} nor ensures the correctness of the converse \cite{blanchini2015switched}. For example, consider the system \eqref{eq4} with two subsystems and the matrices:
		\begin{equation*}
		A_1= \begin{bmatrix}
		1/2 & 1/2 \\ 
		1/3 & 1/2
		\end{bmatrix},\quad
		A_2 = \begin{bmatrix}
		1/2 & 1/2 \\ 
		2/27 & 2/3
		\end{bmatrix}.
		\end{equation*}
		Hence, both $A_1$ and $A_2$ are nonnegative and Schur stable. 
		It is simply seen that no LCPLF can be found, but the dual system has LCPLF $V(x(k)) = \begin{bmatrix}
		4& 3 \end{bmatrix} x(k) $.
	\end{pf}
	
		\subsection{Controller synthesis for the positive T--S fuzzy system}
		%
		%%%------->The goal of this subsection is to derive novel stabilization conditions for nonlinear positive systems.
		In this subsection, an approach based on the LCPLF and PDC controller is proposed. In this technique, the concept of the dual system is used and the Lyapunov function $V(x(k))=x^T(k) p$ is considered to derive the stabilization condition in the form of new LPs.
	Assuming that the state variables can directly be measured, the main aim is then to design a state feedback controller through the PDC structure, which is defined as follows:
	\begin{equation}
	u(k) = \sum_{i=1}^{r}h_i(\vartheta(k))K_i x(k).\label{eq6}
	\end{equation}
	Then, the closed-loop system can be obtained by Substituting \eqref{eq6} into \eqref{eq4}:
	%Substituting \eqref{eq6} into \eqref{eq4} results in the closed-loop system as:
	\begin{align}
	S_c: \left\{ \begin{array}{c}
	x(k+1)=\sum\limits_{i=1}^{r}\sum\limits_{j=1}^{r}h_i(\vartheta(k))h_j(\vartheta(k)) (A_i+B_iK_j)x(k), \\
	\ \ \ z(k)=\sum\limits_{i=1}^{r}\sum\limits_{j=1}^{r}h_i(\vartheta(k))h_j(\vartheta(k))(C_i+D_iK_j)x(k).
	\end{array}   \right. 
	\label{eq7}
	\end{align}
	\
	%The following theorem presents the stability conditions using LP approach.
	
	\begin{thm} \label{thmm4.1}
		For the positive fuzzy system \eqref{eq4}, the state-feedback controller \eqref{eq6} exists such that the closed-loop system \eqref{eq7} is positive and asymptotically stable if and only if there exist a diagonal matrix $Q \in \mathbb{R}^{n\times n} $ and matrix $M_j \in \mathbb{R}^{m\times n} $ such that the following LP conditions are satisfied:
		\begin{subequations}
			\begin{align}
			&\left[  (A_i-I) Q + B_i M_j \right] \textbf{1}^n \ll 0 \label{eq8a}, \quad for \ \ i,j=1,2,...,r,\\
			&Q \gg 0,\label{eq8b}\\
			& M_j  \leq \leq 0,\label{eq8c} \qquad \ \ for \quad j=1,...,r,\\
			&{{\left[ {{a}^{i}} \right]}_{\,h\,t}}\,{{q}_{t}}\,+\,\sum\limits_{z=1}^{r}{\,{{\left[ {{b}^{i}} \right]}_{\,h\,z}}\,{{\left[ {{m}^{j}} \right]}_{\,z\,t}}}\,\ge \ge 0, \quad for \quad h,t=1,...,n. \label{eq8d}
			\end{align}
		\end{subequations}
		where $A_i \triangleq \left[ a^i\right] _{ht}, \  B_i  \triangleq[b_1^i,...,b_n^i]^T,$
		are given square matrices. Then, the controller gain $K_j$ is computed by:
		\begin{equation}
		K_j=M_j Q^{-1} \in \mathbb{R}^{m \times n}.
		\label{f}
		\end{equation}
	\end{thm}
	
	\begin{pf}
		\underline{Sufficiency}: 
		Assume that conditions \eqref{eq8a}-\eqref{eq8c} hold and $K_j=[k_1^j,...,k_n^j]$ are defined through $k_t^j=m_t^jq_t$ for $j=1,..,r$ and $t=1,...,n$. Then, by replacement the fallowing inequality is obtained:
		\begin{align*}
		& [(A_i-I)Q+B_iK_jQ]\textbf{1}^n\ll 0,\\
		& [(A_i-I)+B_iK_j]Q\textbf{1}^n\ll 0,
		\end{align*}
		By definition $p \triangleq Q \textbf{1}^n$ and by using Proposition \eqref{prop1}, one can conclude that the system \eqref{eq7} is asymptotically stable. Now, the condition \eqref{eq8d} implies that the closed-loop system \eqref{eq7} is positive, since 
		\begin{align*}
		{{\left[ {{a}^{i}} \right]}_{\,h\,t}}\,\,+\,\sum\limits_{z=1}^{r}{\,{{\left[ {{b}^{i}} \right]}_{\,h\,z}}\,{{\left[ {{m}^{j}} \right]}_{\,z\,t}}} q^{-1}_t = (A_i+B_iK_j)_{ht} \,\ge \ge 0.
		\end{align*}
		
		\underline{Necessity}:
		Assume that system \eqref{eq7} is asymptotically stable. 
		%The same reason as the proof of Proposition \ref{prop1} is used,
		In \eqref{eq5b}, $A_i$ is replaced by the closed-loop system matrix $(A_{cl}=A_i + B_iK_j)$. Now, the Lyapunov vector $p$ is defined in the form of:
		\begin{equation}
		p = Q \textbf{1}^n, \,\, Q =diag(q_1,q_2,...,q_n)\gg 0.
		\end{equation}
		%where
		%\begin{equation*}
		%Q =diag(q_1,q_2,...,q_n)\gg 0.
		%\end{equation*}
		%
		Now, inequality \eqref{eq5b} can be rewritten as follows:
		\begin{align}
		(A_i + B_i K_j - I) Q \textbf{1}^n \ll 0,\nonumber\\
		\left[  (A_i-I)Q + B_i K_j Q \right] \textbf{1}^n \ll 0.
		\label{eqqq12}  
		\end{align}
		By definition $ M_j \triangleq K_j Q$, \eqref{eqqq12} is equivalent to \eqref{eq8a}.
		To show that the trajectories remain in the positive orthant for all $k \in \mathbb{R}^n_+$, the following inequality should be satisfied:
		\begin{equation}
		(A_i+B_iK_j)_{ht} \geq \geq 0.
		\label{u}
		\end{equation}
		By post-multiplying \eqref{u} by the matrix
		$Q =diag(q_1,q_2,...,q_n)\gg 0,$
		and using definitions ${{A}_{i}}\triangleq {{\left[ {{a}^{i}} \right]}_{\,ht}}$ and ${{B}_{i}}\triangleq {{\left[ {{b}^{i}} \right]}_{\,ht}}$, the following result is obtained:
		\begin{align}
		\left( A_i + B_i K_j \right) _{ht} Q \geq \geq 0,\nonumber\\
		(A_i Q)_{ht} + (B^iK^j Q)_{ht}\geq \geq 0.
		\label{eqqq14}
		\end{align}
		Now, by using definition
		$\,{{\left( {{K}^{j}}\,Q  \right)}_{\,ht}}\,\triangleq\,{{\left( {{M}^{j}} \right)}_{\,ht}}= \,{{\left[ {{m}^{j}} \right]}_{\,ht}}$, the equation \eqref{eqqq14} can be rewritten as:
		\begin{align}
		{{\left[ {{a}^{i}} \right]}_{\,h\,t}}\,{{q}_{t}}\,+\,\sum\limits_{z=1}^{r}{\,{{\left[ {{b}^{i}} \right]}_{\,h\,z}}\,{{\left[ {{m}^{j}} \right]}_{\,z\,t}}}\,\ge \ge 0,\,\,\,\,\,\,\,\,\,\,h,t=1,...,n.
		\label{eqqq15}
		\end{align}
		Inequality \eqref{eqqq15} is equivalent to \eqref{eq8d}.
		Thus, the proof is complete.\qquad \qquad \qquad \qquad \qquad \qquad  \ \ \ \ $\square $ 
	\end{pf}
	
	\begin{rem}
		The obtained results provide not only necessary and sufficient conditions but also there are simple approaches that can lead to a numerical solution for the problem. In fact, since the functions corresponding to the inequality constraints are all linear, the optimization problem can be considered in the standard LPs framework.
	\end{rem}
	
	\begin{rem}
			From the biological point of view, the asymptotic stability of the system based on negativity of the Lyapunov function derivative in Theorem \eqref{thmm4.1} states that the number of tumor cells is moving towards the equilibrium, which indicates that this state is transferred from the malignant region or macroscopic tumor volume into the region of benign growth or microscopic tumor volume, in which the immune system is able to inhibit the tumor growth. Therefore, the side effects of treatment can be significantly reduced. Moreover, based on the same argument, the number of effector cells of the immune system attains a proper level, in which the body’s immune system would be able to start eliminating cancer cells. In addition, it should be noted that according to inequalities \eqref{eq8d} the positivity of the states is preserved, which guarantees the positivity of the number of tumor and effector cells. This constraint has to be satisfied since from the biological view the number of tumor and effector cells cannot be negative.
	\end{rem}

	\section{Simulation results}
	In this section, a simulation study is performed to show the efficiency of the proposed positive fuzzy controller for interactions of the tumor-immune system under immunotherapy and chemotherapy.
	% as well as to compare the obtained results to the previous works.
	Accordingly, a numerical simulation is carried out through MATLAB software.
	Consider system \eqref{d} with matrices $A_i$ and $B_i$, $i=1,...,8$ as shown below:
	\begin{align*}
	& {{A}_{\,1}}=\left[ \begin{matrix}
	5.01780 & 0  \\
	0 & -0.3740  \\
	\end{matrix} \right],\,\,\,\,\,\,\,\,\,{{B}_{\,1}}=\left[ \begin{matrix}
	-0.100 & 0  \\
	0 & 1  \\
	\end{matrix} \right], \\ 
	& {{A}_{\,2}}=\left[ \begin{matrix}
	5.01780 & 0  \\
	0 & -0.3740  \\
	\end{matrix} \right],\,\,\,\,\,\,\,\,\,{{B}_{\,2}}=\left[ \begin{matrix}
	-1000 & 0  \\
	0 & 1  \\
	\end{matrix} \right], \\ 
	& {{A}_{\,3}}=\left[ \begin{matrix}
	5.01780 & 0  \\
	0 & -8.3121  \\
	\end{matrix} \right],\,\,\,\,\,\,\,\,\,\,{{B}_{\,3}}=\left[ \begin{matrix}
	-0.100 & 0  \\
	0 & 1  \\
	\end{matrix} \right], \\ 
	& {{A}_{\,4}}=\left[ \begin{matrix}
	5.01780 & 0  \\
	0 & -8.3121  \\
	\end{matrix} \right],\,\,\,\,\,\,\,\,\,\,{{B}_{\,4}}=\left[ \begin{matrix}
	-1000 & 0  \\
	0 & 1  \\
	\end{matrix} \right], \\ 
	& {{A}_{\,5}}=\left[ \begin{matrix}
	-5.1391 & 0  \\
	0 & -0.3740  \\
	\end{matrix} \right],\,\,\,\,\,\,\,\,\,{{B}_{\,5}}=\left[ \begin{matrix}
	-0.100 & 0  \\
	0 & 1  \\
	\end{matrix} \right], \\ 
	\end{align*}
	\begin{align*}
	& {{A}_{\,6}}=\left[ \begin{matrix}
	-5.1391 & 0  \\
	0 & -0.3740  \\
	\end{matrix} \right],\,\,\,\,\,\,\,\,\,{{B}_{\,6}}=\left[ \begin{matrix}
	-1000 & 0  \\
	0 & 1  \\
	\end{matrix} \right], \\ 
	& {{A}_{\,7}}=\left[ \begin{matrix}
	-5.1391 & 0  \\
	0 & -8.3121  \\
	\end{matrix}\, \right],\,\,\,\,\,\,\,\,{{B}_{\,7}}=\left[ \begin{matrix}
	-0.100 & 0  \\
	0 & 1  \\
	\end{matrix} \right], \\ 
	& {{A}_{\,8}}=\left[ \begin{matrix}
	-5.1391 & 0  \\
	0 & -8.3121  \\
	\end{matrix}\, \right],\,\,\,\,\,\,\,\,{{B}_{\,8}}=\left[ \begin{matrix}
	-1000 & 0  \\
	0 & 1  \\
	\end{matrix} \right].
	\end{align*}
	As it can be seen, matrices $A_i$ and $B_i$ are not positive, however, the matrix $C= I_2$ is common for all the configurations.
	%
	%It is worth noting that for the augmented system, a controller is designed which ensures that states always remain positive while the error is sign free so that as to prevent the progression of cancer.
	%
	A controller system is designed for the augmented system to prevent the progression of cancer, while ensuring the positiveness of state variables.
	Accordingly, the conditions of Theorem \eqref{thmm4.1} are applied. By solving the set of LPs in Eq. \eqref{eq8a}-\eqref{eq8d}, the state-feedback controller gain, and the closed-loop matrices are achieved as follows:
	
	\begin{align*}
	& {{K}_{1}}=\left[ \,\begin{matrix}
	\,0.0329 & 0.0000  \\
	5.0758 & -1.5236  \\
	\end{matrix}\, \right],\,\,\,\,\,\,\,\,\,\,\,\,{{K}_{2}}=\left[ \begin{matrix}
	-0.0018 & 0.0000  \\
	5.3526 & -1.4356  \\
	\end{matrix} \right] \\ 
	& {{K}_{3}}=\left[ \,\begin{matrix}
	\,0.0010 & 0.0000  \\
	5.3542 & -1.4356  \\
	\end{matrix}\, \right],\,\,\,\,\,\,\,\,\,\,\,\,{{K}_{4}}=\left[ \,\begin{matrix}
	\,0.0010 & 0.0000  \\
	5.3377 & -1.4356  \\
	\end{matrix}\, \right] \\
%	\end{align*}
%	\begin{align*}
	& {{K}_{5}}=\left[ \begin{matrix}
	-0.0047 & 0.0000  \\
	5.3460 & -1.4356  \\
	\end{matrix} \right],\,\,\,\,\,\,\,\,\,\,\,{{K}_{6}}=\left[ \,\begin{matrix}
	\,0.0022 & 0.0000  \\
	5.3493 & -1.4356  \\
	\end{matrix}\, \right] \\ 
	& {{K}_{7}}=\left[ \,\begin{matrix}
	\,0.0011 & 0.0000  \\
	5.4053 & -1.4777  \\
	\end{matrix}\, \right],\,\,\,\,\,\,\,\,\,\,\,\,{{K}_{8}}=\left[ \,\begin{matrix}
	\,0.0455 & 0.0000  \\
	5.4152 & -1.4777  \\
	\end{matrix}\, \right] \\ 
	& {{A}_{1}}=\left[ \begin{matrix}
	1.6512 & 0.0000  \\
	0.4982 & 0.8137  \\
	\end{matrix} \right],\,\,\,\,\,\,\,\,\,\,\,\,\,\,\,\,\,\,\,\,{{A}_{2}}=\left[ \begin{matrix}
	1.8790 & 0.0000  \\
	0.5254 & 0.8224  \\
	\end{matrix} \right] \\ 
	& {{A}_{3}}=\left[ \begin{matrix}
	1.6516 & 0.0000  \\
	0.3636 & 0.3380  \\
	\end{matrix} \right],\,\,\,\,\,\,\,\,\,\,\,\,\,\,\,\,\,\,\,{{A}_{4}}=\,\,\left[ \begin{matrix}
	1.5179 & 0.0000  \\
	0.3625 & 0.3380  \\
	\end{matrix} \right] \\ 
	& {{A}_{5}}=\left[ \begin{matrix}
	0.5982 & 0.0000  \\
	0.5247 & 0.8224  \\
	\end{matrix} \right],\,\,\,\,\,\,\,\,\,\,\,\,\,\,\,\,\,\,\,\,{{A}_{6}}=\left[ \begin{matrix}
	0.4291 & 0.0000  \\
	0.5250 & 0.8224  \\
	\end{matrix} \right] \\ 
	& {{A}_{7}}=\left[ \begin{matrix}
	0.5981 & 0.0000  \\
	0.3671 & 0.3352  \\
	\end{matrix} \right],\,\,\,\,\,\,\,\,\,\,\,\,\,\,\,\,\,\,\,\,{{A}_{8}}=\left[ \begin{matrix}
	1.6900 & 0.0000  \\
	0.3677 & 0.3352  \\
	\end{matrix} \right]
	\end{align*}
	
\begin{figure}[t!]
		\begin{center}
			\includegraphics[scale = 0.5]{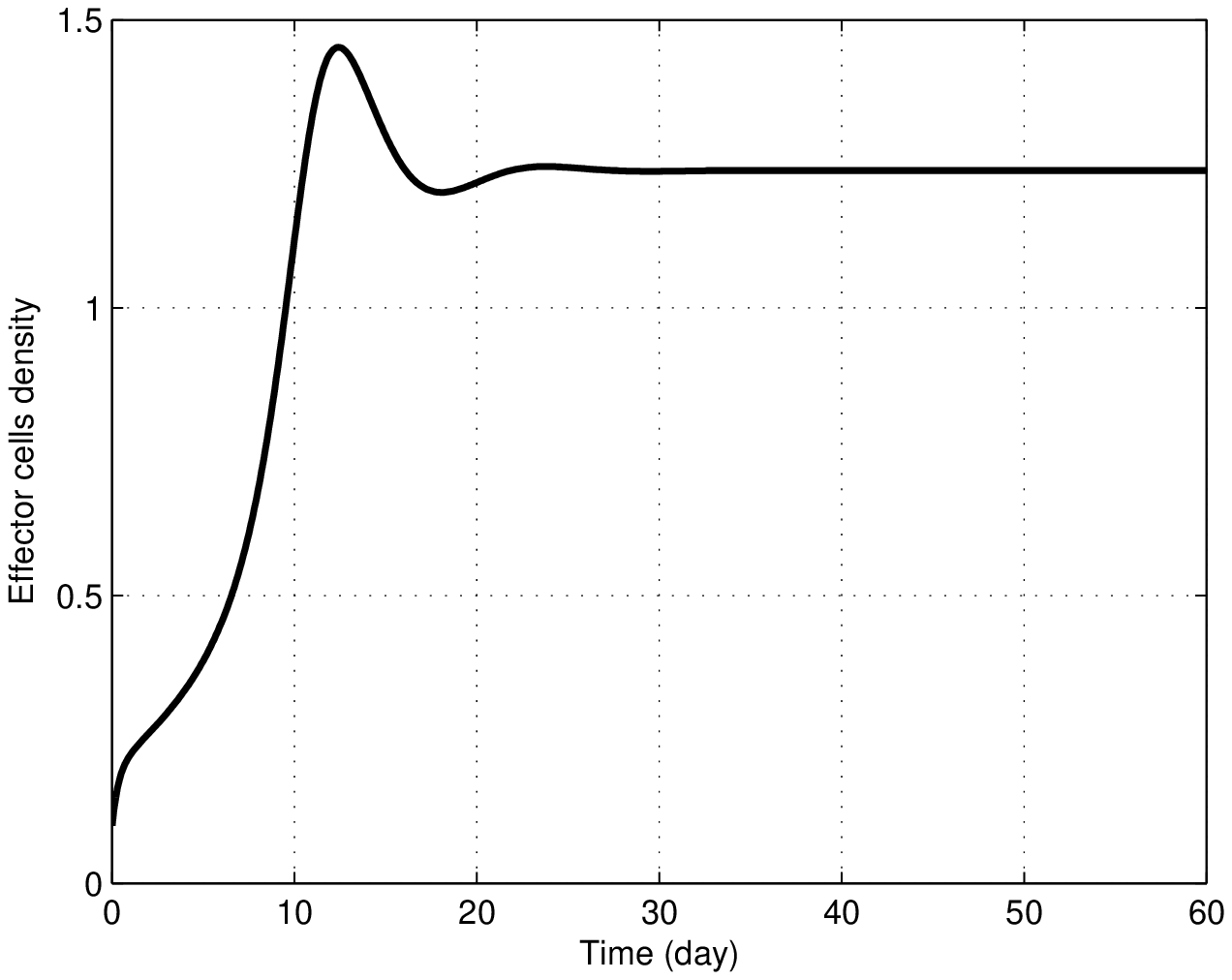}
			\includegraphics[scale = 0.5]{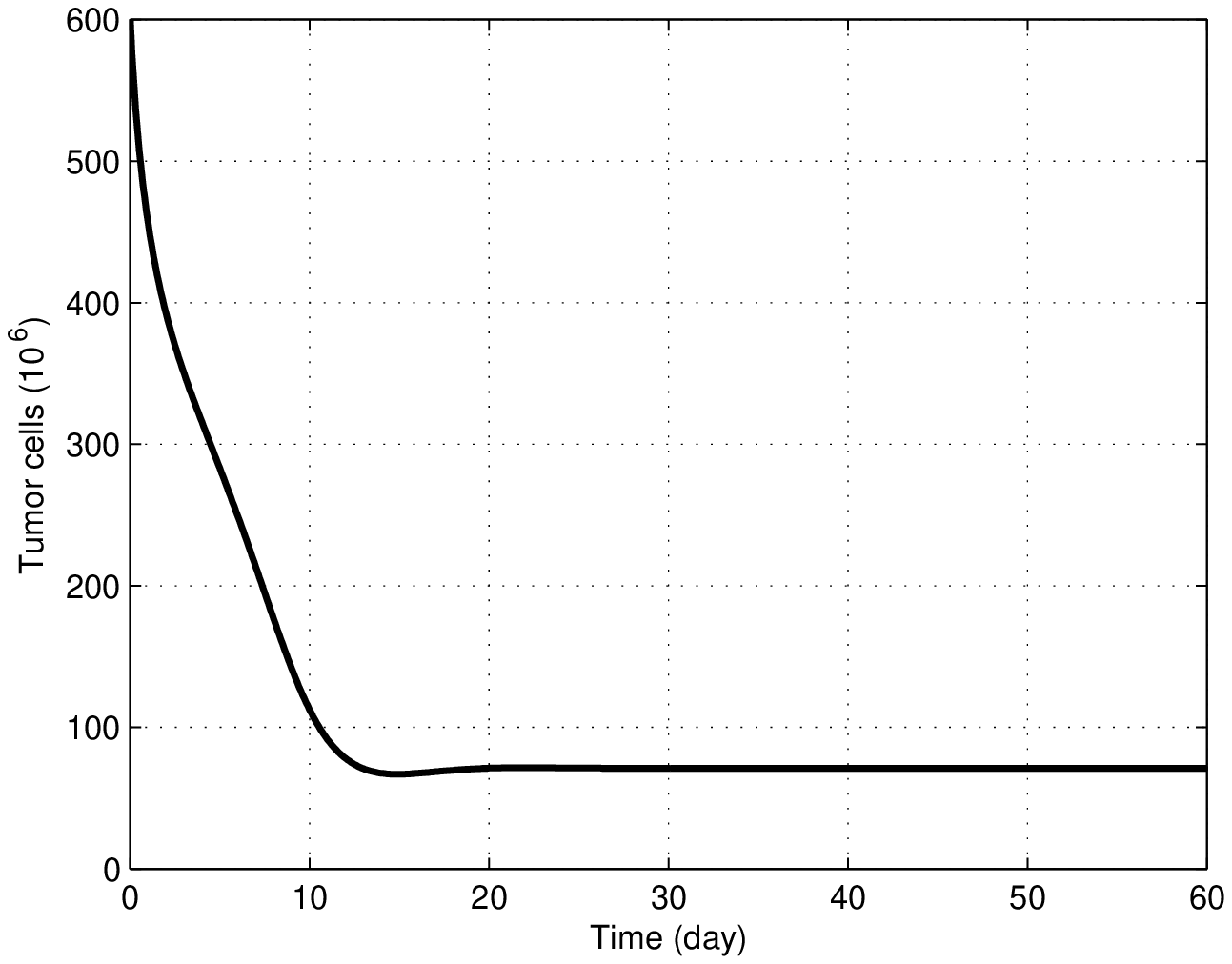}
			\includegraphics[scale = 0.5]{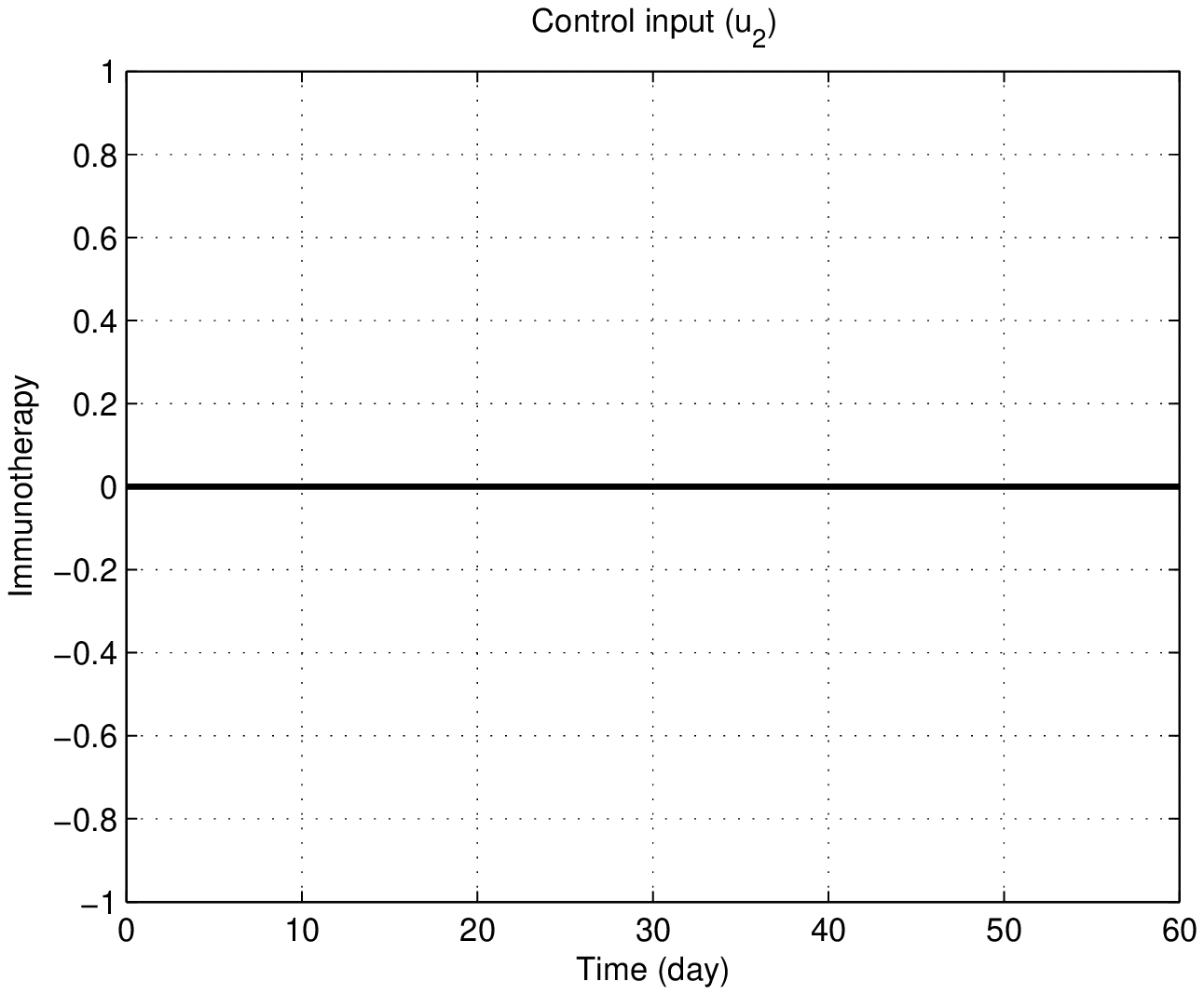}
			\includegraphics[scale = 0.5]{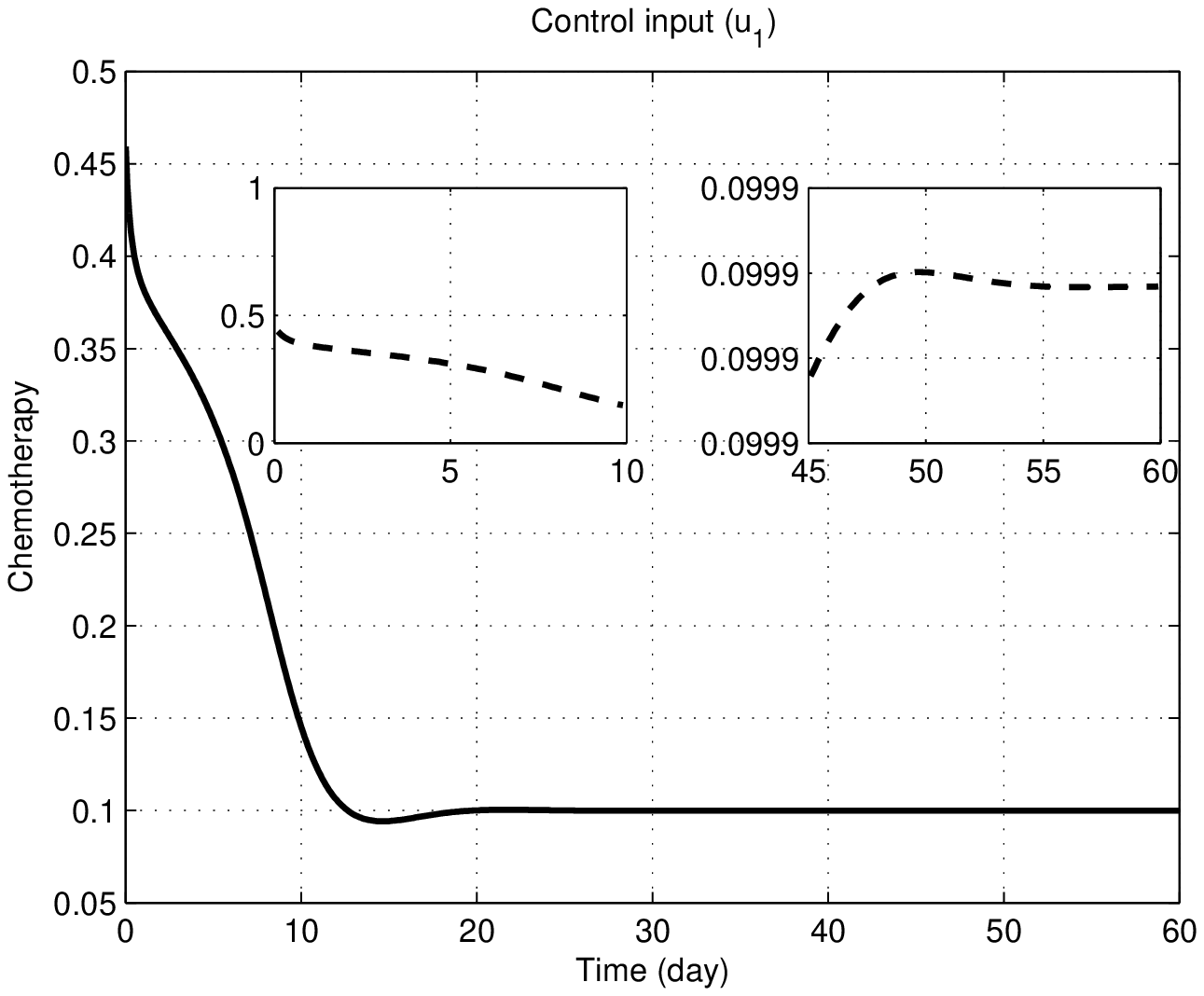}
			\caption{The effect of chemotherapy ($u_1$).}
			\label{fig3}
		\end{center}
	\end{figure}

\begin{figure}[h!]
		\begin{center}
			\includegraphics[scale = 0.5]{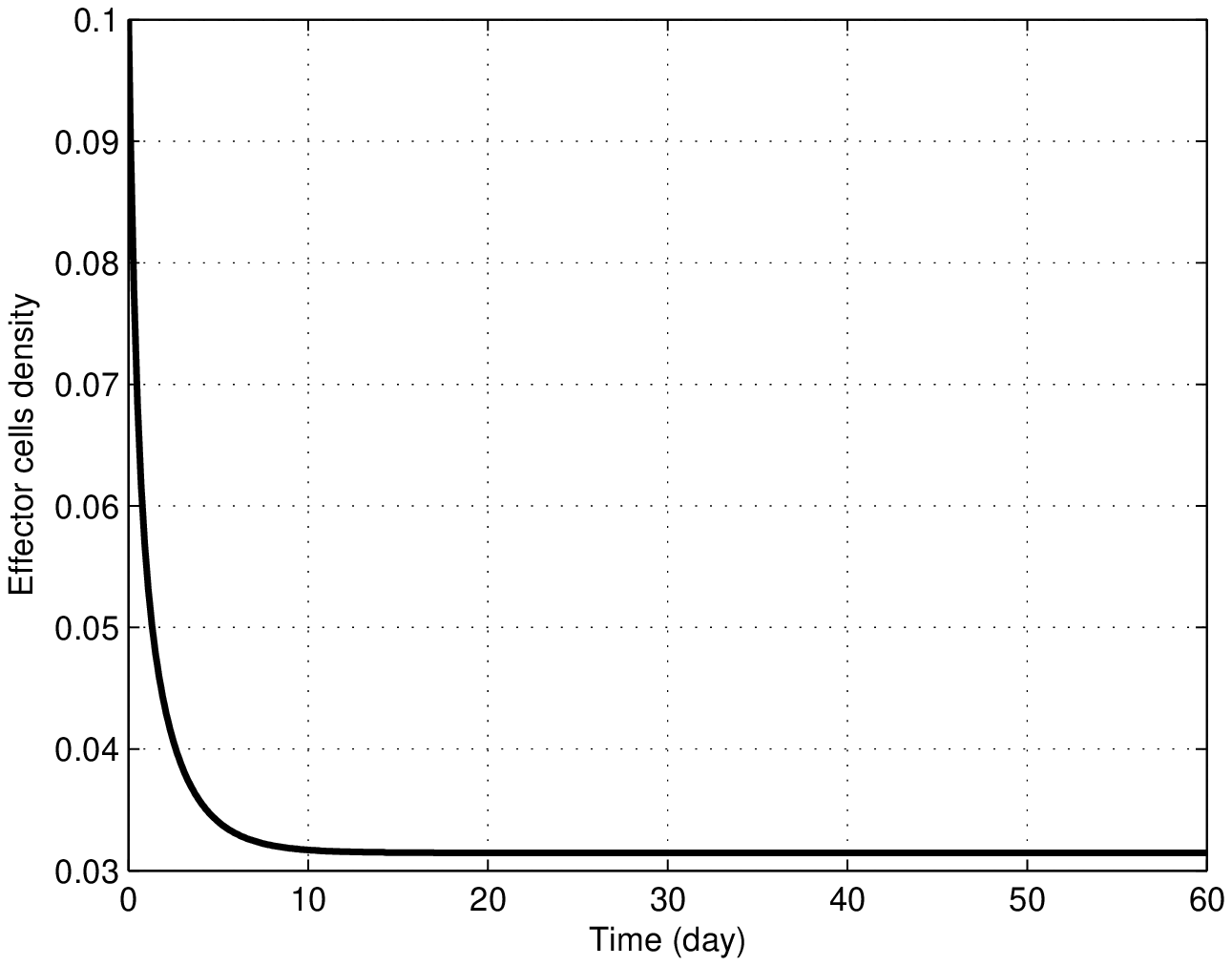}
			\includegraphics[scale = 0.5]{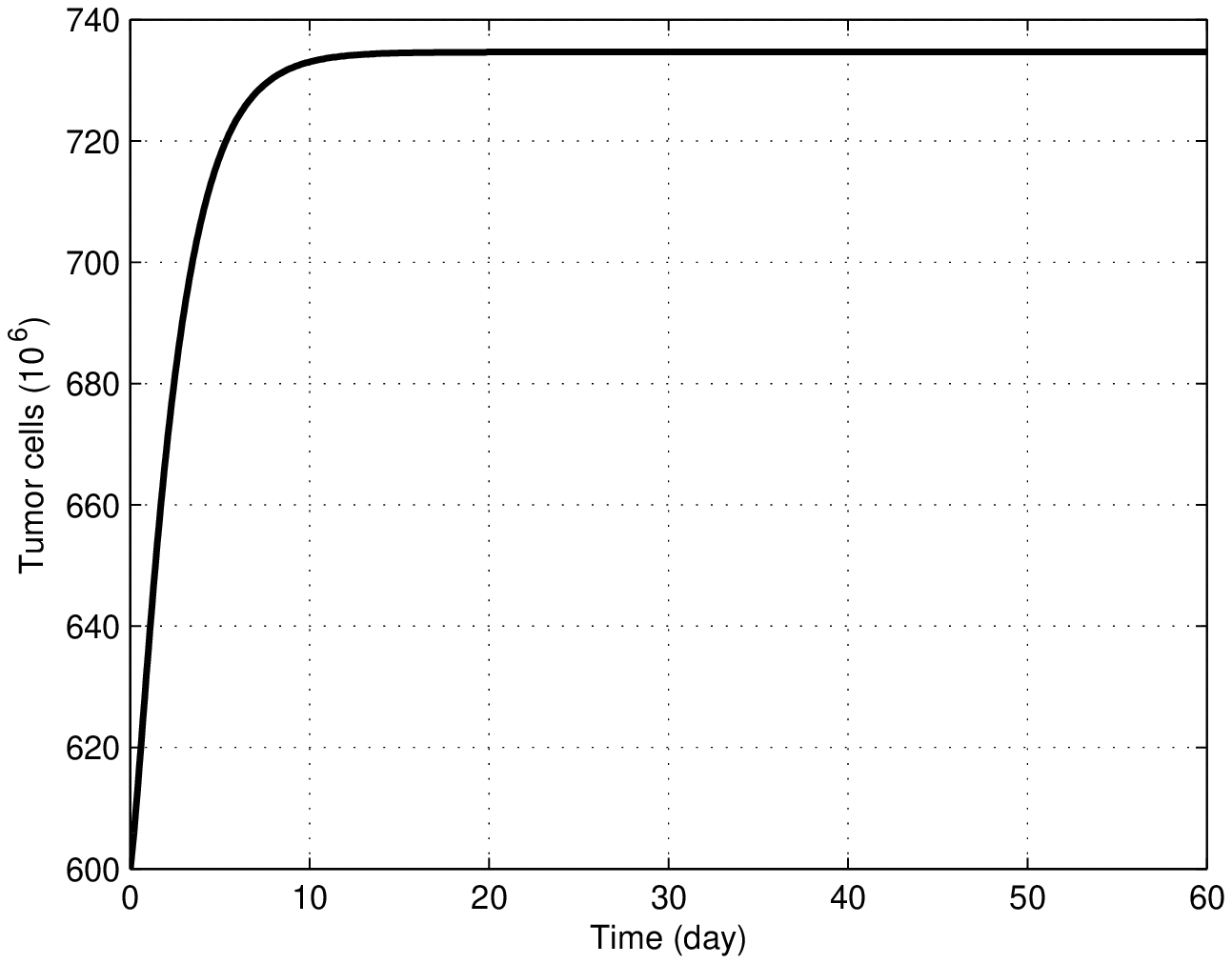}
			\includegraphics[scale = 0.5]{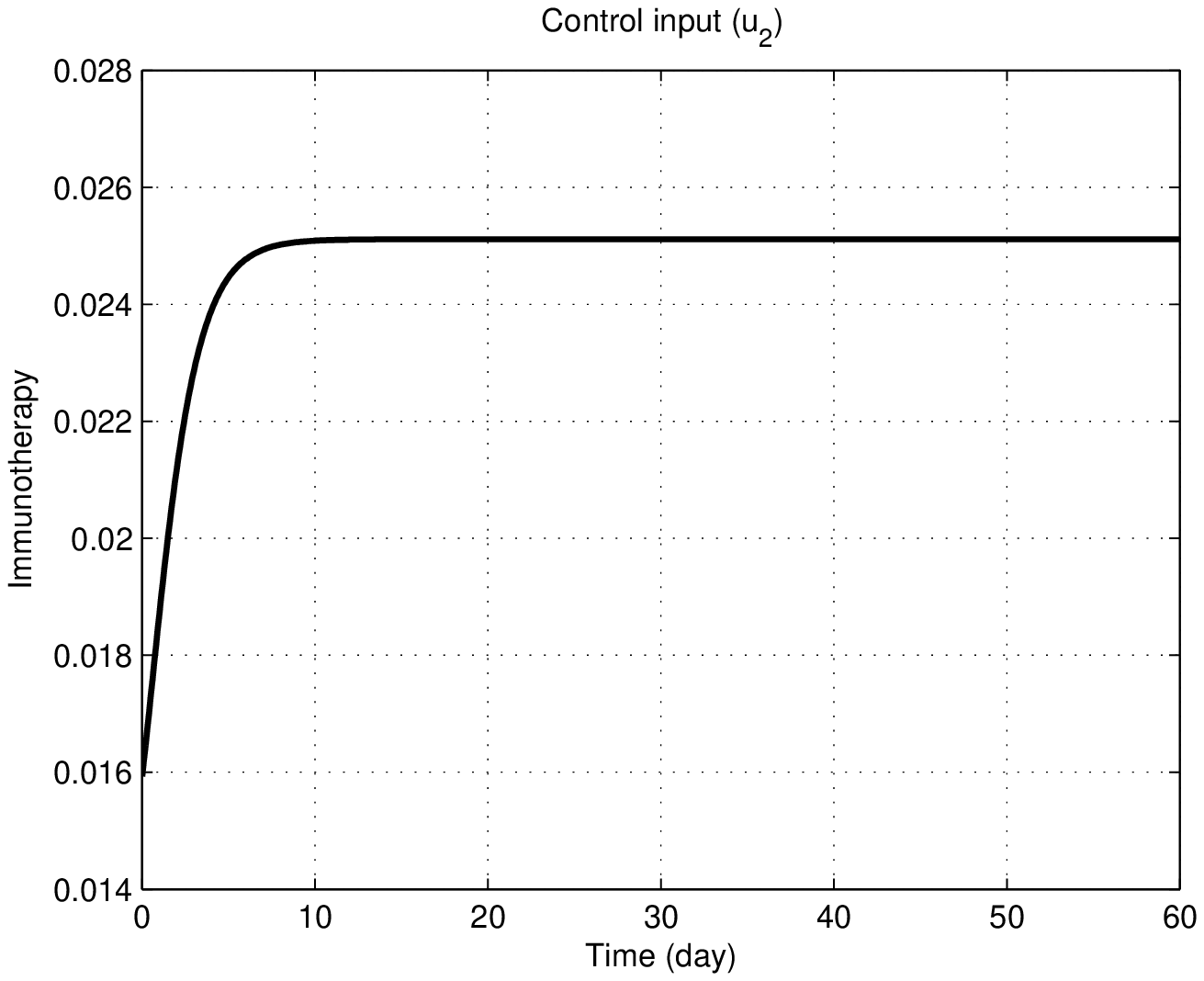}
			\includegraphics[scale = 0.5]{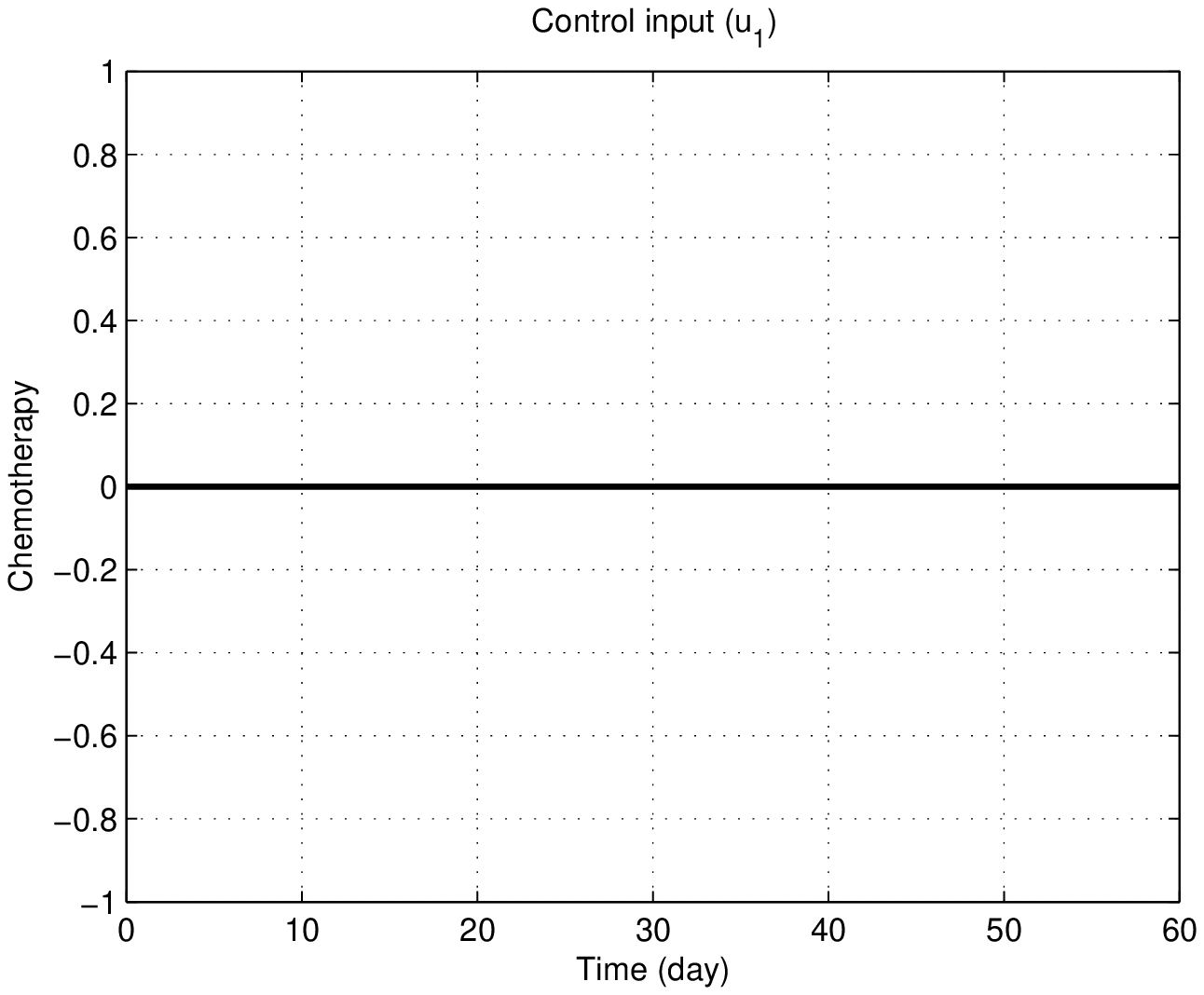}
			\caption{The effect of immunotherapy ($u_2$).}
			\label{fig4}
		\end{center}
	\end{figure}

\begin{figure}[h!]
		\begin{center}
			\includegraphics[scale = 0.5]{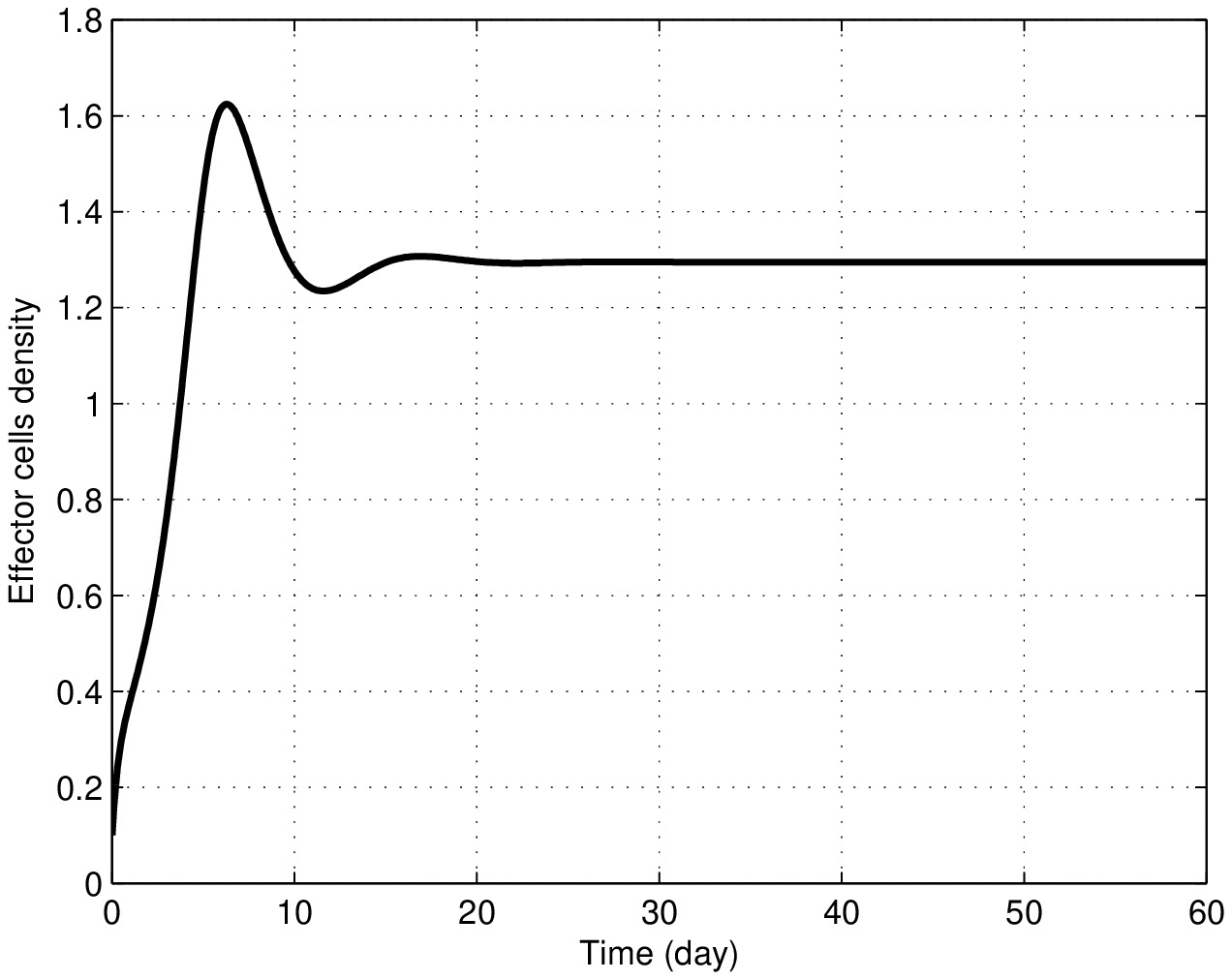}
			\includegraphics[scale = 0.5]{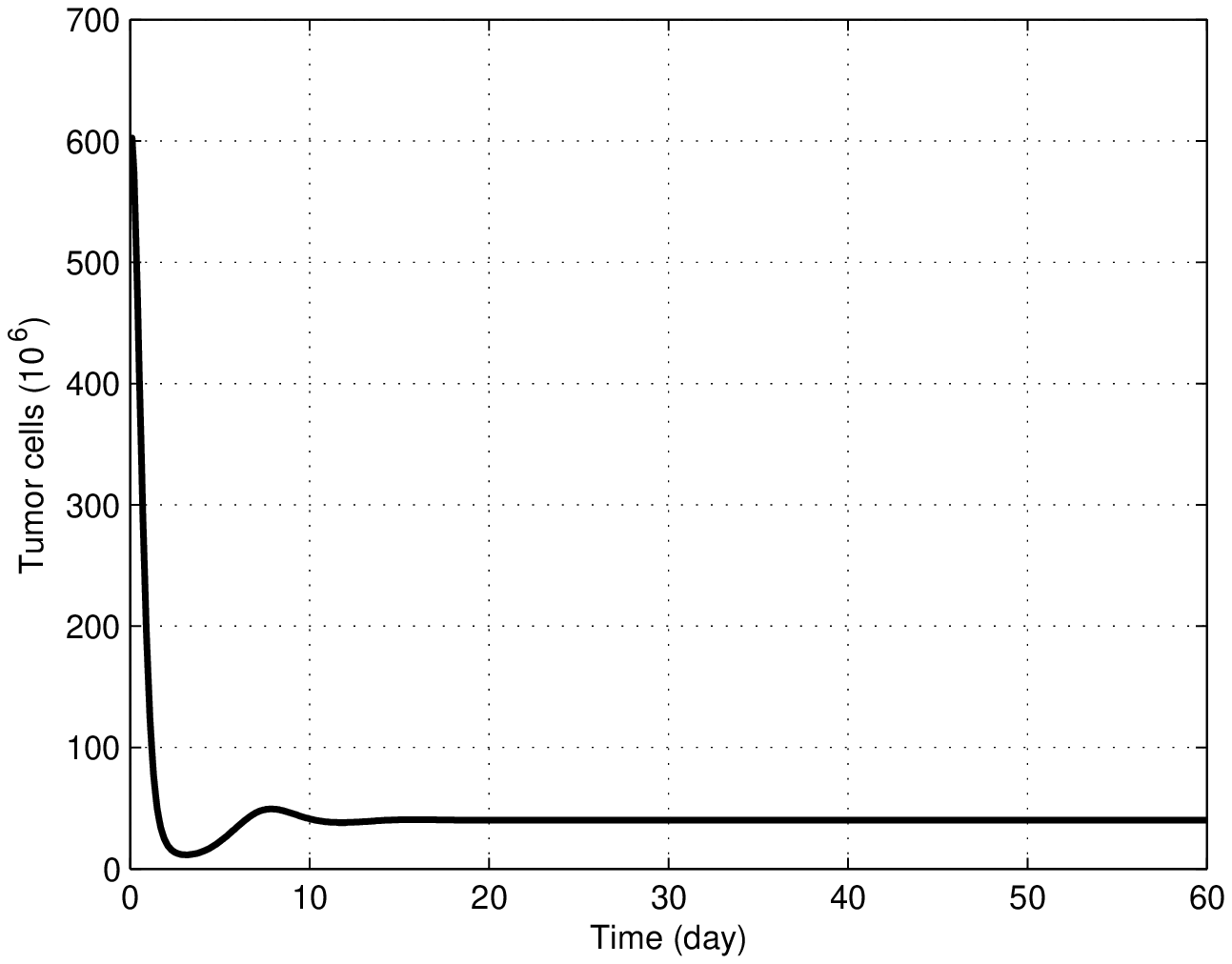}
			\includegraphics[scale = 0.5]{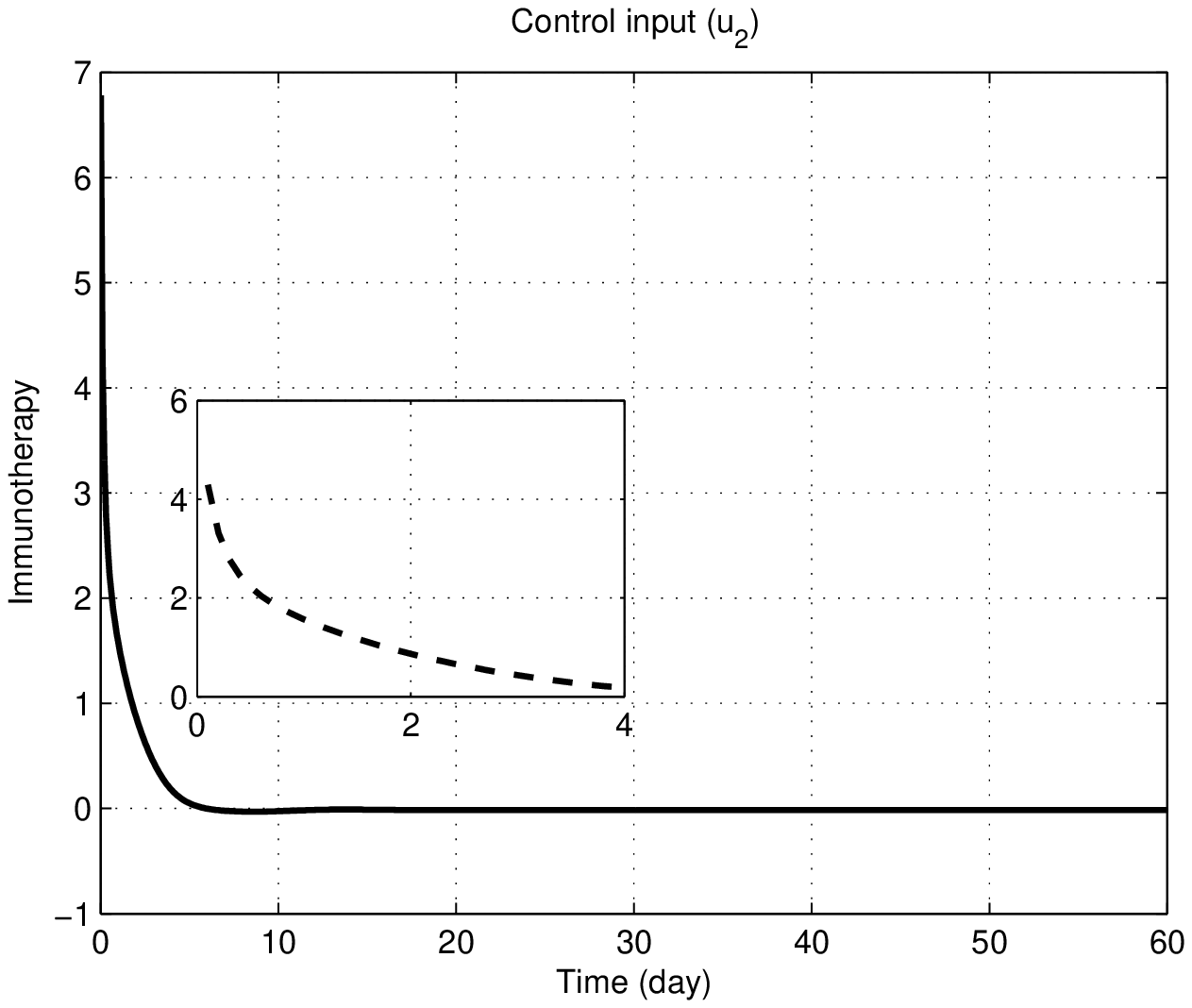}
			\includegraphics[scale = 0.5]{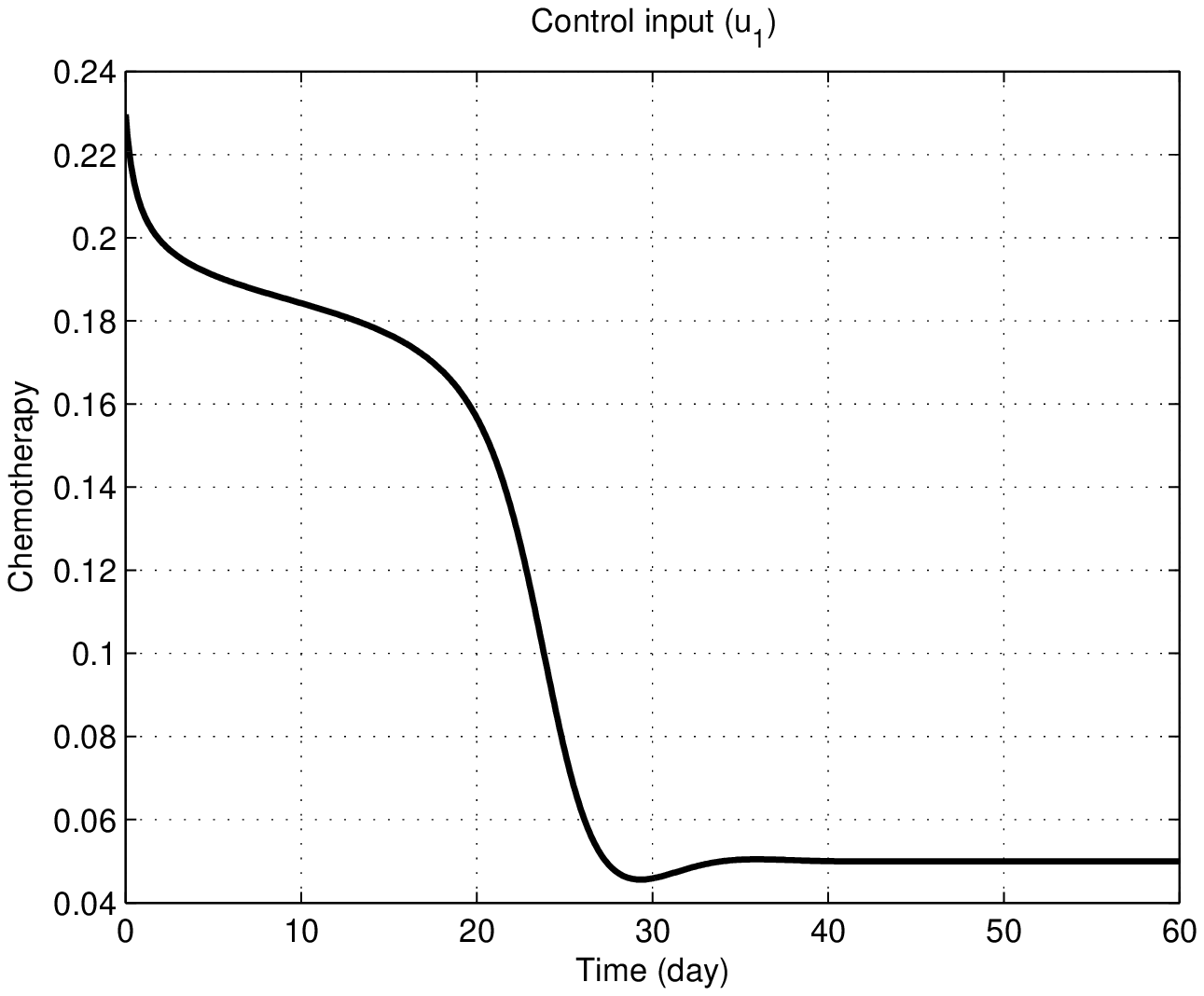}
			\caption{The effect of chemotherapy and immunotherapy. }
			\label{fig5}
		\end{center}
	\end{figure}

	The trajectories of state $x_1$ and $x_2$ and the corresponding control inputs are depicted in Figs. \ref{fig3}-\ref{fig5} for a 60 days of the treatment when the proposed positive fuzzy control is applied to the system.
	The reference and initial conditions are selected within $z_r = \begin{bmatrix}
	50 & 1.6
	\end{bmatrix} ^ T $ and  $ X_0 = \begin{bmatrix}
	600 & 0.1
	\end{bmatrix} ^ T $.
	
	Figure \ref{fig3} depicts the effect of chemotherapy in the cancer treatment. Although the cancer treatment under exclusive chemotherapy has the ability to control the growth of cancer cells, the immediate alteration of chemotherapy dose rate is necessary for the purpose of treatment.
	Figure \ref{fig4} exclusively illustrates the effect of immunotherapy in treating cancer. There is evidently a failed cancer treatment.
	%It is clearly seen that the singular immunotherapy
	In fact, in cases where the initial condition is located in the malignant region, the immune system can not prevent the progression of cancer, and, thus, the tumor growths uncontrollably.
	A combination therapy that is fundamental to a successful treatment, under mixed immunotherapy and chemotherapy is demonstrated in fig. \ref{fig5}. As observed by applying the suggested controller, the tumor load reduces and converges to the equilibrium in the benign region and the number of effector cells attains a proper level after only 20 days upon starting the immunotherapy and chemotherapy treatments.
	%%%that the immunotherapy and chemotherapy are applied.
	These results are a major success in recovering the control approaches for the cancer therapy.
	
	Compared to the results of Ref. \cite{ledzewicz2013optimal}
	that was accomplished on the same model used in this article, with the similar initial condition $ X_0 = \begin{bmatrix}
	600 & 0.1
	\end{bmatrix} ^ T $, it is seen that the designed controller in the current work is able to eliminate more tumor cells and also guarantees the positiveness of the closed-loop system in such a way that no positive constraints imposed to the variables. 
	Additionally, the amplitude of the control inputs in this approach is lower than the one in \cite{ledzewicz2013optimal}. In general, a maximum of 20 \% of the full dose rate of the drug is approximately needed to control the progression of cancer. Accordingly, less control efforts are required for the cancer treatment and, consequently, the side effects of the treatment are reduced significantly.
	%%%---->It is worth mentioning that
	It should be noted that the maximum amount of cancer cells in the proposed method is much lower than the one in \cite{ledzewicz2013optimal}, which indicates a reduction in the likelihood of death during the course of treatment.
	%%%In the other words, in the initial condition $ X_0 = \begin{bmatrix}
	%%%600 & 0.1
	%%%\end{bmatrix} ^ T $ that belongs to the malignant status of the tumor, the controller provided in this paper has been well suited to the goal of treatment.
	In other words, by considering the initial condition $ X_0 = \begin{bmatrix}
	600 & 0.1
	\end{bmatrix} ^ T $ that belongs to the malignant status of the tumor, the proposed controller is an efficient tool for the goal of treatment.  
	
	%
	%\textcolor{red}{As can be seen, the proposed therapeutic approach has decreased the system to the area with less cancer cells, while the cell density of the immune system is at an acceptable level. 
	%In the other words, in the initial condition $X=[600,0.1]^T$ that belongs to the malignant status of cancer, the controller provided in this article has been well suited to the goal of treatment.}

	\section{Conclusion}
	This paper suggests a new fuzzy modeling approach for the nonlinear modified Stepanova’s model, which is one of the efficient and applicable models for the cancer treatment.
	%
	%modest and best
	%
	%the modified Stepanova’s model which is one of the modest and best applicable models was utilized in this work.
	In spite of the previous works that used optimal control strategies, this paper proposes a positive T--S fuzzy controller for the successful treatment of cancer under mixed immunotherapy and chemotherapy. This is based on a new framework that incorporates, for the first time, T--S fuzzy model, linear co-positive Lyapunov function, and parallel distributed compensation via a convex optimization approach.
	%In spite of the previous works that used optimal control strategy for cancer treatment, in this paper a positive T--S fuzzy control strategy is proposed.
	This new framework is a lot invaluable since there are inherent nonnegative sign features for the state variables and their corresponding control inputs.
	The simulation results show that the control structure can be successfully exploited to the tumor-immune system, which leads to the best performance in the sense that it significantly minimizes the volume of the tumor cells and, consequently, reduces the doses of the consumed drug.
	However, the population of the immune competent cells can reach an appropriate level, so that the immune system can prevent tumor growth. 
	Future research will concern an extension of this method to the design of a controller considering uncertainties, disturbance, and unknown factors in the nonlinear model.
	The proposed methodology can also be applied in treating other specific types of cancer.
	
		\appendix
		\section{A review on fuzzy systems and positive systems}
		In this section, brief reviews of T--S fuzzy systems and positive systems are provided. Moreover, several observations and results associated with the positive T--S fuzzy systems that are used throughout this paper are introduced.
	%\subsection{A review on fuzzy systems and positive systems}}
	%
	%This section briefly introduces T--S fuzzy models and positive systems. However, several observations and results associated to positive T--S fuzzy systems are introduced which will be utilized through-out this paper.
	%
	The T--S fuzzy system is one of the most widely used approaches for the stability analysis and controller design of nonlinear systems  \cite{tanaka2004fuzzy}.
	Explicitly, the T--S fuzzy model is described by fuzzy IF-THEN rules, where each locally characterizes linear input-output relations of the system through the sector nonlinearity technique.
	%In this approach, the behavior of the nonlinear system is exactly expressed by the T--S model using the sector nonlinearity technique.
	
	Assume the $i$th plant rule of the T--S model is as follows:\\
	Plant rule $i$:
	IF\;$ \vartheta_1(k) \; is \;M_{i1}\; and ,..., and \; \vartheta_n(k)\; is \;M_{in}\;$, THEN
	\begin{align}
		\left\lbrace\begin{array}{l}
		x(k+1) = A_i x(k) + B_i u(k),\\ 
		z(k) = C_i x(k) + D_i u(k),\\
		x(0) = x_0\succeq 0,
		\end{array}  \right. 
		\label{eq3}
		\end{align} 
	where $x(k)\in \mathbb{R}^n$, $u(k) \in \mathbb{R}^m$, and $z(k)\in \mathbb{R}^l$ are the state vector, input vector, and output vector, respectively. $i\in \mho_r=\{1,...,r\}$, $r$ is the number of IF-THEN rules. $\vartheta_1(k),...,\vartheta_n(k)$ and $M_{ij}$ are the premise variables and the membership functions, respectively. Then, the overall T--S model is given by:
	\begin{align}
	 S: \left\lbrace\begin{array}{l}
		x(k+1) = \sum\limits_{i=1}^{r}h_i(\vartheta (k))(A_i x(k) + B_i u(k),\\ 
		z(k)=\sum\limits_{i=1}^{r}h_i(\vartheta (k))(C_i x(k) + D_i u(k),\\
		x(0) = x_0\succeq 0,
		\end{array}  \right. 
		\label{eq4}
		\end{align} 
	where
	\begin{align*}
	h_i(\vartheta(k))=\dfrac{\omega_i(\vartheta(k))}{\sum_{i=1}^{r}\omega_i (\vartheta(k))},&\qquad \omega_i(\vartheta(k))=\prod_{j=1}^{\mu}M_{ij}(\vartheta_j(k)),\\
	\sum_{i=1}^{r}h_i(\vartheta(k))=1,&\qquad 0\leq h_i(\vartheta(k)) \leq1,
	\end{align*}\\
	and $M_{ij}(\vartheta_j(k))$ is the grade of membership of $\vartheta_j(k)$ in $M_{ij}$.\\
	
	Here, the essential definitions and lemma are introduced.
	
	\begin{defn} \cite{farina2011positive}:
		Given any positive initial condition $x(0)\geq\geq0$ and input $u(k) \geq\geq0 $, the discrete-time system \eqref{eq4} is said to be positive if $x(k)\geq \geq0$ and $z(k)\geq \geq0$ for all integers $ k\in R^n_+$.
	\end{defn}
	
	\begin{defn}  \cite{farina2011positive}:
		In the continuous-time system, a real matrix $A$ is called a Metzler matrix if its off-diagonal elements are positive, i.e. $A_{ij} \geq0$, $i \neq j$.
	\end{defn}
	
	\begin{defn} \cite{farina2011positive}: 
		In the discrete-time system, a real matrix $\it A$ is called Schur if all its eigenvalues are strictly inside the unit circle.
	\end{defn}
	
	\begin{defn} A linear map $V(x(k))=p^T x(k)$ with $V(0)=0$ is said to be an LCPLF for positive system $x(k+1)=A x(k)$, if the following conditions hold for all $x\in \mathbb{R}^n_+ $:
		\begin{equation*}
		\begin{cases}
		V(x(k))>0, \\
		\Delta V(x(k))<0.
		\end{cases}
		\end{equation*}
	\end{defn}

	\begin{lem} \cite{benzaouia2016advanced}: \label{lemma1}
		The discrete-time T--S fuzzy system \eqref{eq4} is positive if and only if
		\begin{align*}
		A_i\geq \geq0,\ B_i\geq \geq0,\ C_i\geq \geq0,\  D_i\geq \geq0,
		\end{align*}
		where $i=1,...,r.$
	\end{lem}
	
	\begin{rem} 
		%%% -----> It is worth noting that,
		When $x(0) \in R^n_+ $ is not hold, $x(k) \geq \geq 0$, may not satisfy for all the sets of nonnegative integers even if all the conditions of Lemma \eqref{lemma1} hold.
		In order to certify the advisable efficiency of Lemma \eqref{lemma1},
		the satisfaction of $x(0) \in R^n_+ $ is assumed.
	\end{rem}
	
	%\section*{References}
	\bibliography{mybibfile}
	%\bibliographystyle{ieeetr-fa}
	%\bibliography{myliba}
	
\end{document}